\documentclass[pra,aps,twocolumn,amsmath,amssymb,superscriptaddress]{revtex4-1}
\usepackage{epsfig,amsmath}
\usepackage{subfigure}
\usepackage{graphicx}% Include figure files
\usepackage{dcolumn}% Align table columns on decimal point
\usepackage{stmaryrd}
\usepackage{mathrsfs}
\usepackage{pifont}
\usepackage{amsthm}
\usepackage{braket}
\usepackage{amssymb}
\usepackage{bm}
\usepackage{latexsym}
\usepackage[colorlinks=true,linkcolor=blue,citecolor=blue]{hyperref}
\usepackage{color}
\usepackage{epstopdf}
\usepackage[ruled]{algorithm2e}

\begin{document}

\title{Achieving the Heisenberg limit of metrology via measurement on an ancillary qubit}

\author{Peng Chen}
\affiliation{School of Physics, Zhejiang University, Hangzhou 310027, Zhejiang, China}

\author{Jun Jing}
\email{Contact author: jingjun@zju.edu.cn}
\affiliation{School of Physics, Zhejiang University, Hangzhou 310027, Zhejiang, China}

\begin{abstract}
In the scenario of the probe-ancilla interaction, we propose a quantum metrology protocol by the unconditional measurement on the ancillary qubit after an optimized period of joint evolution from product state. Its key element is the construction of two parallel evolution paths by the measurement that can transform the probe system (a spin ensemble) from an eigenstate of a collective angular momentum operator $|j,m\rangle$ to a superposed state $(|j,m\rangle+|j,-m\rangle)/\sqrt2$. With synchronous parametric encoding and qubit measurement, the quantum Fisher information about the phase encoded in the probe system with optimized initial states can exactly attain the Heisenberg scaling $N^2$ with respect to the probe size (spin number) $N$. The quadratic scaling behavior is not sensitive to the imprecise control over the joint evolution time, the time delay between encoding and measurement, and the coherence in the probe ensemble or the ancillary system that would be degraded by local dephasing. The classical Fisher information of the spin ensemble is found to saturate with its quantum counterpart, irrespective of the idle joint evolution after the parametric encoding. We suggest that both Greenberger-Horne-Zeilinger (GHZ)-like states and nonlinear Hamiltonian are {\em not} necessary resources for exceeding the standard quantum limit in metrology precision since in our protocol even thermal states can hold an asymptotic quadratic scaling.
\end{abstract}

\maketitle

\section{Introduction}
Quantum metrology aims to push the measurement precision to the ultimate limit constrained by the intrinsic uncertainty of quantum mechanics~\cite{Sun2010fisher,ma2011quantum,genoni2012optical,escher2012quantum,
zhong2013fisher,giovannetti2004quantum}. A fundamental limit of quantum metrology in uncorrelated systems is the standard quantum limit (SQL) on the errors of parameter estimation, which scales linearly as $1/\sqrt{N}$ with $N$ the total number of measurements or system components. This shot-noise scaling law is essentially rooted in the central limit theorem of classical statistics~\cite{kay1993fundamentals}. If nonclassical resources, e.g., the Greenberger-Horne-Zeilinger (GHZ) state in atomic systems and the NOON state in photonic systems, are introduced to the system, the sensitivity of quantum measurement can be enhanced and the precision limit of parameter estimation can exceed the SQL and approach the Heisenberg limit (HL), representing a scaling law inverse to the number of probe units with $1/N$. Achieving this sensitivity with a large quantum number is an ongoing target of a wide range of potential applications including atomic clocks~\cite{Ludlow2015optical,Katori2011optical}, gravitational wave detection~\cite{Caves1981quantum}, biological sensing~\cite{Taylor2016quantum,Mauranyapin2017evanescent}, and magnetometry~\cite{Jones2009magnetic}.

Standard metrology protocols in multiparticle systems rely on GHZ-like and squeezed states. However, generating a large GHZ-like state is a highly nontrivial task. One of the generation strategies is to apply the entangling gates to qubits~\cite{song2019generation,choi2014optimal,barends2014superconducting,kaufmann2017scalable}, but the achievement is very limited. So far an $18$-qubit GHZ state with fidelity $\sim0.525$ has been generated on a quantum processor~\cite{song2019generation}. Dynamical process driven by the one-axis twisting (OAT) Hamiltonian $H_{\rm OAT}=\chi J_z^2$~\cite{kitagawa1993squeezed} provides another viable method for generating GHZ-like states from separable spin-coherent states. For example, a GHZ-like state was generated on the electronic spin $J=8$ of dysprosium atoms, where the OAT Hamiltonian results from the spin-dependent energy shifts~\cite{chalopin2018quantum}. $\chi$ is yet so weak that the desired evolution time $\chi t=\pi/2$ is quite long, setting obstacles for state generation due to decoherence and particle loss~\cite{sorensen2001many,pezze2009entanglement,agarwal1997atomic,leibfried2005creation}. Measurement and postselection can reduce the evolution time at the cost of the success probability~\cite{alexander2020generating}. Squeezed spin states are typically generated through the nonlinear interactions~\cite{wineland1992spin,kitagawa1993squeezed}. Collective OAT interaction, which is popular in Bose-Einstein condensations (BECs)~\cite{gross2010nonlinear,riedel2010atom}, trapped ions~\cite{bohnet2016quantum,lu2019global}, and superconducting qubits~\cite{song2019generation,xu2020probing}, can give rise to a sub-HL noise reduction $\propto1/N^{2/3}$ for $N$ particles with the optimized duration $\chi t\simeq3^{1/6}N^{-2/3}$~\cite{wineland1992spin,kitagawa1993squeezed}. Under the two-axis twisting (TAT) interaction $H_{\rm TAT}=\chi(J_z^2-J_y^2)$, it was theoretically claimed that the squeezing degree can approach Heisenberg limit $\propto1/N$ with the optimized time $\chi t\simeq\ln(4N)/(2N)$~\cite{zhang2017cavity,borregaard2017one}. But realizing $H_{\rm TAT}$ remains as a challenge in the current platforms~\cite{helmerson2001creating,borregaard2017one,macri2020spin}.

The interactions between the probes and the additional dimensions from an ancillary system can also increase the precision of parametric estimation to surpass SQL without entanglement~\cite{boixo2007generalized,demkowicz2014using,zhang2022approaching,fan2024achieving,chen2024qubit}. In a protocol for measuring the frequency of identical probe units~\cite{fan2024achieving}, the information of the parameter can be obtained through the measurements on the ancillary qubit by tracing out the degree of freedom of the probe system, where the quantum Fisher information (QFI) can be simplified in the Bloch representation~\cite{zhong2013fisher}. The estimation precision of the parameter in the probe Hamiltonian can achieve the Heisenberg scaling in terms of both the evolution time and the number of probe units by tailoring the interaction Hamiltonian, finding proper time points for measurements, and tuning the interaction strength.

In the scenario of probe-ancilla interaction, we propose a metrology protocol by tracing out the ancillary system to estimate the phase parameter $\theta$ imprinted to the probe system (a spin ensemble) during a fast rotation. Starting from the product state of probe and ancilla, the probe can evolve from a collective angular momentum eigenstate $|j,m\rangle$ to a superposed state $(|j,m\rangle+|j,-m\rangle)/\sqrt2$ after an optimal period of joint evolution and  the unconditional measurement on qubit. If the initial probe state is a polarized state $|j,j\rangle$, then an ideal GHZ state can be generated. If the to-be-estimated parameter $\theta$ is encoded in the probe state at the same time of measurement, the quantum Fisher information about $\theta$ can approach the Heisenberg scaling $N^2$ in terms of the total spin number $N$. It is interesting to find that an asymptotic Heisenberg-scaling behavior about the metrology precision holds when the probe or the ancillary qubit is under the influence of local dephasing or even prepared as a thermal state. And this scaling behavior is not sensitive to the imprecise control over the joint evolution time and the time delay between the parametric encoding and the measurement on the ancillary qubit. Moreover, the classical Fisher information (CFI) of our protocol is found to be coincident with QFI by the projective measurements on the probe system, irrespective of the idle evolution after all the operations.

The rest of this work is structured as follows. In Sec.~\ref{roleFisherInf} we briefly recall the classical Fisher information and its quantum counterpart, distinguishing the critical role of probe state played in the estimation theory. In Sec.~\ref{metrologyProtocol} we describe the circuit model of our metrology protocol assisted by the unconditional measurement on the ancillary qubit. In Sec.~\ref{prepareGHZ} we investigate the parametric conditions of our protocol for attaining the Heisenberg limit in metrology precision. We analyze the quantum Fisher information under the influence of noise on the probe and ancillary qubit in Secs.~\ref{QFI coherence probe} and \ref{QFI coherence qubit}, respectively. In Sec.~\ref{ImperfectControl} we discuss the sensitivity of our metrology protocol to the imprecise controls over the evolution time and the time delay between encoding and measurement on qubit. In Sec.~\ref{ClassicalFisherInf} we calculate CFI in our protocol as the amount of extractable information from the probability distribution of the output state of the probe system upon projective measurements. The entire work is summarized in Sec.~\ref{conclusion}. In the Appendix, we present QFI out of our protocol simulated with real experimental parameters, which is bounded by environmental dissipation.

\section{Essential Role of probe state}\label{roleFisherInf}

The Fisher information and the Cram\'{e}r-Rao bound are foundational concepts in quantifying the limits of precision in parameter estimation. These two concepts are pivotal in both classical and quantum domains, offering insight to the optimal accuracy achievable in the situation free of noise and deviation.

In the classical domain, the Fisher information provides a measure about the sensitivity of the probability distribution of the system state to the changes in parameters. Assuming that $\{p(x_i|\theta),\theta\in\mathbb{R}\}_{i=1}^d$ is the probability density conditioned on the fixed value of the phase parameter $\theta$ with the measurement outcomes $\{x_i\}$, the classical Fisher information $F_C$ is defined as~\cite{cramer1999mathematical}
\begin{equation}\label{CFI}
F_C=\sum_{i=1}^d\frac{[\partial_\theta p(x_i|\theta)]^2}{p(x_i|\theta)},
\end{equation}
where $\partial_\theta\equiv\partial/\partial\theta$. When the observable is a continuous variable, the summation in Eq.~(\ref{CFI}) should be replaced by an integral.

The quantum analog of the Fisher information is formally generalized from Eq.~(\ref{CFI}) and defined as
\begin{equation}\label{QFI original}
F_Q={\rm Tr}\left(\rho_\theta L_\theta^2\right)
\end{equation}
in terms of the symmetric logarithmic derivative (SLD) operator $L_{\theta}$, which is a Hermitian operator defined as
\begin{equation}
\partial_\theta\rho_\theta=\frac{1}{2}\left(\rho_\theta L_\theta+L_\theta\rho_\theta\right).
\end{equation}
By diagonalizing the density matrix as
\begin{equation}\label{state after encoding}
\rho_\theta=\sum_{i=1}^dp_i|\psi_i(\theta)\rangle\langle\psi_i(\theta)|,
\end{equation}
with $p_i\geq0$ and $\sum_{i=1}^dp_i=1$, the elements of the SLD operator are completely defined under the condition $p_i+p_j\neq0$. Therefore, Eq.~(\ref{QFI original}) can be expressed as~\cite{braunstein1994statistical}
\begin{equation}\label{QFI after encoding}
\begin{aligned}
F_Q=&\sum_{i=1}^d4p_i\langle\partial_\theta\psi_i(\theta)|\partial_\theta\psi_i(\theta)\rangle\\
-&\sum_{i,j=1}^d\frac{8p_ip_j}{p_i+p_j}|\langle\psi_i(\theta)|\partial_\theta\psi_j(\theta)\rangle|^2.
\end{aligned}
\end{equation}

Consider a general unitary parametrization process
\begin{equation}\label{unitary}
U_\theta^\dagger U_\theta=\mathcal{I}^d,
\end{equation}
with $\mathcal{I}^d$ being the identity matrix of $d$ dimensions, then the density matrix in Eq.~(\ref{state after encoding}) becomes
\begin{equation}\label{state before encoding}
\rho_\theta=U_\theta\rho_0U_\theta^\dagger,
\end{equation}
where
\begin{equation}\label{initial spectral}
\rho_0=\sum_{i=1}^dp_i|\psi_i\rangle\langle\psi_i|
\end{equation}
is the spectral decomposition of the initial state before parametric encoding. By Eq.~(\ref{state before encoding}), Eq.~(\ref{QFI after encoding}) reduces to~\cite{braunstein1994statistical,braunstein1996generalized,zhang2013quantum,liu2014quantum}
\begin{equation}\label{QFI}
F_Q=\sum_{i=1}^d4p_i\langle\psi_i|\mathcal{H}^2|\psi_i\rangle-\sum_{i,j=1}^d\frac{8p_ip_j}{p_i+p_j}|\langle\psi_i|\mathcal H|\psi_j\rangle|^2
\end{equation}
with the effective phase generator $\mathcal{H}\equiv iU_\theta^\dagger(\partial_\theta U_\theta)$. Note that even if the parametrization process $U_\theta$ is nonunitary~\cite{brody2012mixed}, Eq.~(\ref{QFI}) still holds if the initial state is a pure state or its spectral decomposition satisfies the condition
\begin{equation}\label{orthogonal condition}
\langle\psi_i|U_\theta^\dagger U_\theta|\psi_j\rangle=0
\end{equation}
for all $i\neq j$.

For a pure initial state $\rho_0=|\psi\rangle\langle\psi|$, the quantum Fisher information in Eq.~(\ref{QFI}) reduces to
\begin{equation}\label{purestate}
F_Q=4\langle\psi|\mathcal{H}^2|\psi\rangle-4|\langle\psi|\mathcal{H}|\psi\rangle|^2.
\end{equation}
One can find that the variance of the phase generator $\mathcal{H}$ becomes maximized when the initial state $|\psi\rangle $ is an equally weighted superposition of the eigenvectors $|\lambda_{\rm max}\rangle$ and $|\lambda_{\rm min}\rangle$ with $\lambda_{\rm max}$ and $\lambda_{\rm min}$ the maximum and minimum eigenvalues of $\mathcal{H}$, respectively~\cite{giovannetti2006quantum,pang2014quantum}, i.e.,
\begin{equation}\label{psi}
|\psi\rangle=\frac{1}{\sqrt 2}(|\lambda_{\rm max}\rangle+e^{i\phi}|\lambda_{\rm min}\rangle),
\end{equation}
where $\phi$ is an arbitrary relative phase. Thus the maximal QFI is~\cite{pang2014quantum}
\begin{equation}
F_Q^{\rm max}=(\lambda_{\rm max}-\lambda_{\rm min})^2.
\end{equation}
An important observation is that $|\psi\rangle$ can be a maximal entangled state for a many-body or continuous-variable system, e.g., GHZ state and NOON state. For a parametric encoding about the $x$-axis, i.e., $\mathcal{H}=J_x$ of a high-spin system, the relevant GHZ state is
\begin{equation}\label{GHZ state}
|{\rm GHZ}\rangle=\frac{1}{\sqrt 2}(\mathcal I^{N+1}+e^{-i\phi}e^{i\pi J_z})|j,j\rangle_x,
\end{equation}
where $|j,m\rangle_x$ is the eigenstate of the collective spin operators $J_x$ with eigenvalue $m$, $-j\leq m\leq j$, and $N=2j$. It suggests that the superposed state in Eq.~(\ref{psi}) with a maximally departed distribution in eigenspace can be generated by constructing two parallel evolution paths. In this work, this idea is carried out through the measurement on the ancillary qubit, by which a polarized state can be transformed to be a GHZ state.

\section{Metrology with ancillary qubit}\label{metrologyProtocol}

In this paper, we consider a metrology model consisting of a high-spin probe (spin ensemble) and an ancillary spin-$1/2$. The full Hamiltonian is based on the so-called ZZZZ model~\cite{braun2018quantum} including the free part $H_0$ and the interaction part $H_I$ ($\hbar\equiv1$), i.e.,
\begin{equation}\label{H}
H=H_0+H_I=\omega_PJ_z+\omega_A\sigma_z+gJ_z\sigma_z,
\end{equation}
where $J_\alpha=\sum_{k=1}^N\sigma_\alpha^k/2$, $\alpha=x,y,z$, denotes the collective spin operator with $\sigma_{\alpha}^k$ the $\alpha$-component of Pauli matrix of the $k$th probe spin, $\omega_P$ and $\omega_A$ represent the energy splitting of the probe spin and the ancillary spin, respectively, and $g$ is the coupling strength between the two components.

The interaction Hamiltonian in Eq.~(\ref{H}) is feasible in various experiments. In the nitrogen-vacancy (NV) centers~\cite{xie2021beating}, $J_z$ and $\sigma_z$ describe the $^{13}$C nuclear spins and the NV electron spin, respectively. The magnetic field parallel to the quantization axis of the NV center sets an additional splitting between $m_s=\pm1$ states, and allows us to isolate a two-level subsystem comprising $m_s=0,1$. In the hybrid qubit-photon-magnon systems, the Hamiltonian can be distilled from a dispersive interaction between the Kittel mode of the magnon and a superconducting qubit through a microwave cavity~\cite{lachance2020entanglement}. The collective spin operators of magnon $J_{\alpha}$, $\alpha=x,y,z$, are associated with the bosonic operators of Kittel mode through the Holstein-Primakoff (HP) transformation. In addition, our Hamiltonian can be realized by two cavities containing two-component BECs coupled by cavity quantum electrodynamics~\cite{pellizzari1995decoherence,pyrkov2013entanglement}, where the operators $J_z$ and $\sigma_z$ correspond to the Schwinger boson operators $a^\dagger a-b^\dagger b$ in the relevant cavity modes with $a^\dagger$ and $b^\dagger$ the bosonic creation operators for two orthogonal quantum states.

\begin{figure}[htbp]
\begin{centering}
\includegraphics[width=0.9\linewidth]{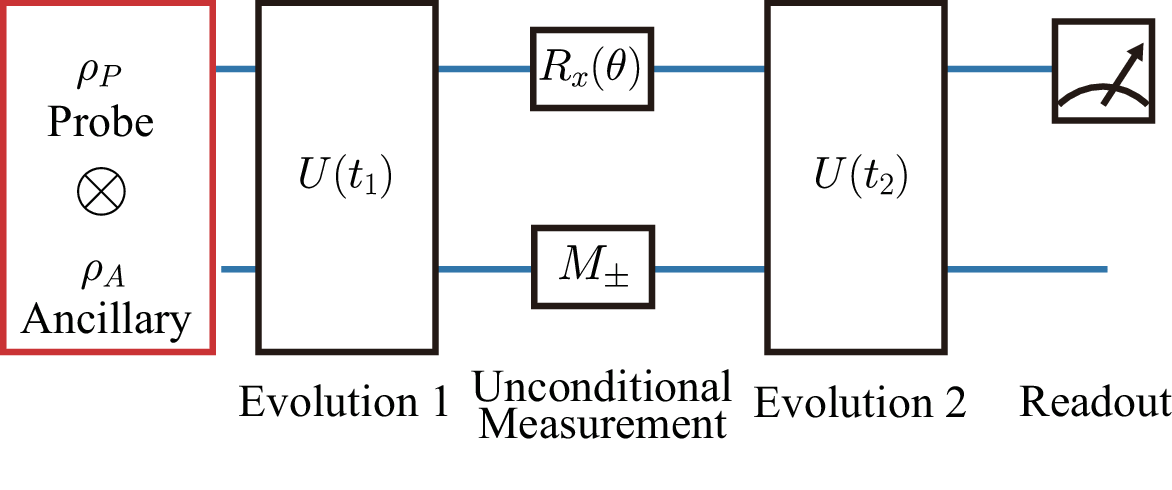}
\caption{Circuit model of our measurement-based metrology protocol. Evolution $1$ and $2$ denote the free joint unitary evolution $U(t_1)$ and $U(t_2)$, respectively, between which the unconditional measurement $M_\pm$ in the basis of $\sigma_x$ is performed on the ancillary qubit and meanwhile a to-be-estimated parameter $\theta$ is encoded in the probe state via a unitary rotation $R_x(\theta)$ about the $J_x$ direction. The output state is determined by projective measurements on the probe system.}\label{protocol}
\end{centering}
\end{figure}

The probe spin ensemble and the ancillary qubit are assumed to be initially separable, i.e., the input state of the full system is a tensor product state $\rho_P\otimes\rho_A$ as indicated by Fig.~\ref{protocol}. The whole evolution operator of the circuit can be described by
\begin{equation}\label{U}
\begin{aligned}
U_{\theta,\pm}&=U(t_2)R_x(\theta)M_\pm U(t_1)\\& =e^{-iHt_2}e^{-i\theta J_x}|\pm \rangle \langle \pm|e^{-iHt_1},
\end{aligned}
\end{equation}
where $|\pm\rangle=(|e\rangle\pm|g\rangle)/\sqrt 2$ is the eigenvector of $\sigma_x$. Here $|g\rangle$ and $|e\rangle$ denote the ground and excited states of the ancillary spin, respectively. During Evolution $1$ and Evolution $2$, the probe system and the ancillary qubit experience a joint evolution lasting $t_1$ and $t_2$, respectively. On the stage of Unconditional Measurement, the projection $M_\pm=|\pm\rangle\langle\pm|$ is performed on the ancillary qubit. Experimentally, the single-shot measurement of the ancillary qubit state, e.g., electron spin in NV centers, is performed with the collected fluorescence~\cite{xie2021beating}. In the magnon-qubit systems, the state of the ancillary qubit can be measured by virtue of the Jaynes-Cummings nonlinearity of a superconducting cavity coupled to the transmon qubit~\cite{reed2010high}. At the same time, the phase parameter $\theta$ is imprinted to the probe system (a spin ensemble) by a spin rotation $R_x(\theta)=\exp(-i\theta J_x)$ of a negligible duration. Alternatively, the interaction Hamiltonian between the two evolutions can be temporarily switched off. In practice, $\theta$ can be accumulated during the precessing about the $z$ axis~\cite{meyer2001experimental,gross2010nonlinear,ockeloen2013quantum} induced by a certain interaction between the probe and a to-be-measured system. Then the rotation about the $x$ axis could be realized by a sequence of $R_x(\theta)=R_y(\pi/2)R_z(\theta)R_y(-\pi/2)$, where $R_y(\pm\pi/2)$ indicates the $\pi/2$ pulse about $y$ axis. On the last stage of the circuit, projective measurements are performed on the output state about the probe system $\rho_\pm(\theta)\equiv{\rm Tr}_A[U_{\theta,\pm}\rho_P\otimes\rho_AU_{\theta,\pm}^\dagger]$ with ${\rm Tr}_A$ the partial trace over the ancillary qubit, whose probability distribution can be used to infer the classical Fisher information about the estimated parameter $\theta$.

\section{Preparation of a state with maximally departed distribution}\label{prepareGHZ}

The key insight underlying our protocol is to use the free joint unitary evolution for Evolution $1$ and the unconditional measurement $M_\pm$, equivalent to a two-path evolution, to prepare a state with a maximally departed distribution in the probe space. Irrespective of the measurement outcome, the whole process continues and each unraveling of the probe is usually a maximally entangled state when it starts from a pure state. Suppose the probe system and the ancillary qubit are prepared as pure states $\rho_P=|\psi\rangle\langle\psi|$ and $\rho_A=|\varphi\rangle\langle\varphi|$, respectively. After the unitary evolution $U(t_1)$ and a measurement on the ancillary qubit in the $\sigma_x$ basis, the unnormalized output state reads
\begin{equation}\label{unnormalized state}
\begin{aligned}
|\Psi_\pm\rangle=M_\pm U(t_1)|\psi\rangle\otimes|\varphi\rangle=\langle\pm|e^{-iHt_1}|\varphi\rangle|\psi\rangle\otimes|\pm\rangle,
\end{aligned}
\end{equation}
where the subscripts $\pm$ indicate the measurement results. Using Eq.~(\ref{H}), the evolution operator of Evolution $1$ can be expressed as
\begin{equation}
\begin{aligned}
e^{-iHt_1}=&\frac{1}{2}\exp\{-i[\omega_A+(\omega_P+g)J_z]t_1\}\\
&\times[P_+(t_1)\otimes|+\rangle\langle+|+P_-(t_1)\otimes|+\rangle\langle-|\\
&+P_+(t_1)\otimes|-\rangle\langle-|+P_-(t_1)\otimes|-\rangle\langle+|],
\end{aligned}
\end{equation}
with $P_\pm(t)=\mathcal I^{N+1}\pm e^{2i(\omega_A+gJ_z)t}$, where $N=2j$ is the total spin number of the probe ensemble. It is found that the evolution operator can be considered as the sum of two evolution paths $P_\pm$ for the probe, which can be unraveled upon choosing a proper state of the ancillary qubit, i.e.,
\begin{equation}\label{ancillary optimal}
|\varphi\rangle_{\rm opt}=|\pm\rangle.
\end{equation}
Without loss of generality, we first consider the initial state $|\varphi\rangle_{\rm opt}=|+\rangle$. Consequently, the unnormalized output state of the composite system in Eq.~(\ref{unnormalized state}) becomes
\begin{equation}\label{psi pm}
\begin{aligned}
|\Psi_\pm\rangle=&\langle\pm|e^{-iHt_1}|\varphi\rangle_{\rm opt}|\psi\rangle\otimes|\pm\rangle\\
=&\frac{1}{2}e^{-i\omega_At_1}\left(\mathcal I^{N+1}\pm e^{2i\omega_At_1}e^{2igt_1J_z}\right)\\
&e^{-i(\omega_P+g)t_1J_z}|\psi\rangle\otimes|\pm\rangle.
\end{aligned}
\end{equation}
One can find that Eqs.~(\ref{GHZ state}) and (\ref{psi pm}) share the same formation up to a global phase. It indicates that the GHZ state can be generated with a proper initial state of the probe system
\begin{equation}\label{state optimal}
|\psi\rangle_{\rm opt}=|j,\pm j\rangle_{\rm opt}=e^{i(\omega_P+g)t_{1,{\rm opt}}J_z}|j,\pm j\rangle_x,
\end{equation}
where the subscript $x$ denotes the eignestate of $J_x$, and an optimized joint-evolution time for Evolution $1$,
\begin{equation}\label{time optimal}
t_1=t_{1,{\rm opt}}(n_1)=\left(n_1+\frac{1}{2}\right)\frac{\pi}{g}
\end{equation}
with $n_{1}$ integer. Here $|j,m\rangle_{\rm opt}$ with $-j\leq m\leq j$ denote the eigenstates for the optimized collective angular momentum operator
\begin{equation}\label{Jopt_operator}
\begin{aligned}
  &J_{\rm opt}=e^{i(\omega_P+g)t_{1,{\rm opt}}J_z}J_xe^{-i(\omega_P+g)t_{1,{\rm opt}}J_z}\\ =&\cos[(\omega_P+g)t_{1,{\rm opt}}]J_x-\sin[(\omega_P+g)t_{1,{\rm opt}}]J_y,
\end{aligned}
\end{equation}
according to Eq.~(\ref{psi pm}).

Then after the encoding $R_x(\theta)$ and the idle evolution $U(t_2)$ in Eq.~(\ref{U}), the unnormalized output state will become
\begin{equation}\label{output state}
\begin{aligned}
|\Psi_{\theta,\pm}\rangle=&\frac{1}{2}e^{-i\omega_At_{1,{\rm opt}}}U(t_2)R_x(\theta)\\
&(|j,j\rangle_x\pm i^Ne^{2i\omega_At_{1,{\rm opt}}}|j,-j\rangle_x)\otimes|\pm\rangle,
\end{aligned}
\end{equation}
with a probability $\mathcal N_\pm=\langle\Psi_{\theta,\pm}|\Psi_{\theta,\pm}\rangle=1/2$. The probability (normalization factor) is always $50\%$, irrespective of the to-be-determined phase parameter $\theta$. Using the pure-state formula in Eq.~(\ref{purestate}), the effective QFI is found to be
\begin{equation}\label{QFI pm}
\begin{aligned}
F_{Q,\pm,{\rm eff}}&=4\mathcal N_\pm\left[\langle\partial_\theta\Psi_{\theta,\pm }^\prime|\partial_\theta\Psi_{\theta,\pm}^\prime\rangle-|\langle\Psi_{\theta ,\pm }^\prime|\partial_\theta\Psi_{\theta,\pm }^\prime\rangle|^2\right]\\
&=4\langle\partial_\theta\Psi_{\theta,\pm}|\partial_\theta\Psi_{\theta,\pm }\rangle=N^2/2
\end{aligned}
\end{equation}
with the normalized state $|\Psi_{\theta,\pm}^\prime\rangle =|\Psi_{\theta,\pm}\rangle/\sqrt{\mathcal N_\pm}$. Since the unconditional measurement on the ancillary qubit is free from whether the result is $|+\rangle$ or $|-\rangle$, one can use the sum of the effective QFIs $F_{Q,+,{\rm eff}}$ and $F_{Q,-,{\rm eff}}$ to indicate the full QFI, i.e.,
\begin{equation}\label{QFI optimal}
F_Q=F_{Q,+,{\rm eff}}+F_{Q,-,{\rm eff}}=N^2.
\end{equation}
It is therefore found that the Heisenberg scaling can be approached under proper qubit state, optimized probe state, and optimized joint evolution time $t_1$ before encoding and measurement.

When the ancillary qubit is prepared in another optimized state $|\varphi\rangle_{\rm opt}=|-\rangle$, the composite system  becomes
\begin{equation}
\begin{aligned}
|\Psi_\pm\rangle=&\frac{1}{2}e^{-i\omega_At_1}\left(\mathcal I^{N+1}\mp e^{2i\omega_At_1}e^{2igt_1J_z}\right)\\
&e^{-i(\omega_P+g)t_1J_z}|\psi\rangle\otimes|\pm\rangle,
\end{aligned}
\end{equation}
which is similar to Eq.~(\ref{psi pm}). Upon the optimal conditions in Eqs.~(\ref{state optimal}) and (\ref{time optimal}), one can obtain the same result as Eq.~(\ref{QFI optimal}).

\section{Quantum Fisher information under decoherence}\label{QFI coherence}

Our protocol is based on an open-quantum-system-like Hamiltonian~(\ref{H}). To a certain degree of coarse-graining, the ancillary qubit mimics an environment surrounding the central probe system used for quantum metrology.
It is important to address the decoherence channel for either probe or ancilla in practice, which may be detrimental to the metrological performance of quantum measurements. This section is devoted to the estimation over the impact of local dephasing noises on QFI.

\subsection{Effect of probe coherence}\label{QFI coherence probe}

When the ancillary qubit is prepared at the pure state $|\varphi\rangle_{\rm opt}=|\pm\rangle$ and the probe system is in the eigenstate $|\psi\rangle=|j,m\rangle_{\rm opt}$ of the optimized collective angular momentum operator~(\ref{Jopt_operator}), QFI of the output state by our protocol is found to be
\begin{equation}\label{QFI for pure state}
\begin{aligned}
F_Q&=F_{Q,+,{\rm eff}}+F_{Q,-,{\rm eff}}\\
&=4\left[\langle\psi|J_{\rm opt}^2|\psi\rangle-|\langle\psi|e^{-i\pi J_z}J_{\rm opt}|\psi\rangle|^2\right]
\end{aligned}
\end{equation}
using the definition in Eq.~(\ref{purestate}). The operator $e^{-i\pi J_z}$ plays the role of the spin-rotation around the $z$-axis with an angle $\pi$ such that the second term in Eq.~(\ref{QFI for pure state}) vanishes $|\langle j,m|e^{-i\pi J_z}J_{\rm opt}|j,m\rangle_{\rm opt}|^2=m^2|\langle j,m|j,-m\rangle_{\rm opt}|^2=0$. For a superposed state, e.g., $|\psi\rangle=|\psi_m\rangle=a_m|j,m\rangle_{\rm opt}+b_me^{-i\phi_m}|j,-m\rangle_{\rm opt}$ with $a_m$, $b_m$, and $\phi_m$ real numbers and $a_m^2+b_m^2=1$, we have
\begin{equation}
|\langle\psi_m|e^{-i\pi J_z}J_{\rm opt}|\psi_m\rangle|^2=4a^2_mb^2_mm^2\sin^2\phi_m.
\end{equation}
It is then found that the second term vanishes when $\phi_m=n_2\pi$ with an integer $n_2$ and hence $F_Q=4m^2$. For a more general superposed state $|\psi\rangle=\sum_mc_m|\psi_m\rangle$ with $c_m$'s the normalized complex amplitudes, Eq.~(\ref{QFI for pure state}) becomes
\begin{equation}\label{QFI pure probe}
F_Q=4\sum_mm^2|c_m|^2\left(1-4a^2_mb^2_m\sin^2\phi_m\right).
\end{equation}
It is found that $F_Q=N^2$ is attainable if and only if the probe state is prepared at $|j,j\rangle_{\rm opt}$, $|j,-j\rangle_{\rm opt}$, or a superposed state over them, i.e., $a|j,j\rangle_{\rm opt}\pm b|j,-j\rangle_{\rm opt}$ with arbitrary real numbers $a$ and $b$. It implies that the polarized states $|j,j\rangle_{\rm opt}$ and $|j,-j\rangle_{\rm opt}$ can be regarded as resource states for approaching HL in metrology precision besides the standard GHZ-like states. Generally, any polarized state can be transformed to the GHZ state for quantum metrology as long as the relevant angular momentum operator in Eq.~(\ref{Jopt_operator}) and the circuit model in Fig.~\ref{protocol} can be constructed.

If the probe system $\rho_{P}$ is initialized as a mixed state, such as $\rho_0$ in Eq.~(\ref{initial spectral}), then the output state of the composite system after three stages of evolution~(\ref{U}) becomes
\begin{equation}
\rho_{\theta,\pm}=U_{\theta,\pm}^\prime\rho_0U_{\theta,\pm}^{\prime\dagger}\otimes|\pm\rangle\langle\pm|
\end{equation}
under the optimized initial state of the ancillary qubit $|\varphi\rangle_{\rm opt}=|+\rangle$ and the joint-evolution time in Eq.~(\ref{time optimal}). Here the effective evolution operator reads
\begin{equation}
  U_{\theta,\pm}^\prime=e^{-iHt_2}e^{-i\theta J_x}\langle\pm|e^{-iHt_{1,{\rm opt}}}|+\rangle/\sqrt{\mathcal N_\pm^\prime}
\end{equation}
with the normalized factor
\begin{equation}
  \mathcal N_\pm^\prime={\rm Tr}[\langle\pm|e^{-iHt_{1,{\rm opt}}}|+\rangle\rho_0\langle +|e^{iHt_{1,{\rm opt}}}|\pm\rangle].
\end{equation}
The effective evolution operator is {\em not} always unitary unless
\begin{equation}
\begin{aligned}
&U_{\theta,\pm}^{\prime\dagger}U_{\theta,\pm}^\prime=\frac{1}{\mathcal N_\pm^\prime}\langle+|e^{iHt_{1,{\rm opt}}}|\pm\rangle\langle\pm|e^{-iHt_{1,{\rm opt}}}|+\rangle\\
=&\frac{2\mathcal I^{N+1}\pm[e^{2i\omega_At_{1,{\rm opt}}}+(-1)^Ne^{-2i\omega_At_{1,{\rm opt}}}]e^{i\pi J_z}}{2\pm[e^{2i\omega_At_{1,{\rm opt}}}+(-1)^Ne^{-2i\omega_At_{1,{\rm opt}}}]{\rm Tr}[e^{i\pi J_z}\rho_0]}\\
&=\mathcal I^{N+1},
\end{aligned}
\end{equation}
which is equivalent to require
\begin{equation}\label{optimal omegaA}
\omega_A=\frac{N+1+2n_3}{4t_{1,{\rm opt}}(n_1)}\pi
\end{equation}
with $n_3$ integer. Thus the qubit frequency $\omega_A$ has to be dependent on the optimized joint-evolution time $t_{1,{\rm opt}}$ in Eq.~(\ref{time optimal}) or the probe-ancilla coupling strength $g$. In this case, QFI for the mixed initial probe $\rho_0$ can be simply described by Eq.~(\ref{QFI}), i.e.,
\begin{equation}\label{QFI for mixed state}
\begin{aligned}
F_Q&=F_{Q,+,{\rm eff}}+F_{Q,-,{\rm eff}}=\sum_{i=1}^d4p_i\langle\psi_i|J_{\rm opt}^2|\psi_i\rangle\\
&-\sum_{i,j=1}^d\frac{8p_ip_j}{p_i+p_j}|\langle\psi_i|e^{-i\pi J_z}J_{\rm opt}|\psi_j\rangle|^2
\end{aligned}
\end{equation}
after the conditional unitary evolution.

A behavior asymptotic to the Heisenberg scaling can appear even when the second term in Eq.~(\ref{QFI for mixed state}) is only partially eliminated under a mixed state. For example, one can suppose that the probe system is prepared as
\begin{equation}\label{probeMixed}
\begin{aligned}
\rho_P&=a_m^2|j,m\rangle_{\rm opt}\langle j,m|+b_m^2|j,-m\rangle_{\rm opt}\langle j,-m|\\
&+c(e^{-i\phi_m}|j,m\rangle_{\rm opt}\langle j,-m|+{\rm H.c.}),
\end{aligned}
\end{equation}
where $c$ is a positive number with $0\leq c\leq |a_mb_m|$. In the case of a nonvanishing coherence, i.e., $0<c\leq |a_mb_m|$, the probe state in Eq.~(\ref{probeMixed}) can be diagonalized as
\begin{equation}\label{diagonalizeProbe}
\rho_P=c_{m,+}^2|\psi_{m,+}\rangle\langle\psi_{m,+}|+c_{m,-}^2|\psi_{m,-}\rangle\langle\psi_{m,-}|,
\end{equation}
where $c_{m,\pm}^2=(1\pm\Omega)/2$ with $\Omega=\sqrt{4c^2+\Delta^2}$, $\Delta=a_m^2-b_m^2$, and
\begin{equation}
|\psi_{m,\pm}\rangle=\frac{(\Delta\pm\Omega)e^{-i\phi_m}|j,m\rangle_{\rm opt}+2c|j,-m\rangle_{\rm opt}}{\sqrt{4c^2+(\Delta\pm\Omega)^2}}.
\end{equation}
With $|\varphi\rangle_{\rm opt}=|+\rangle$, the optimized joint-evolution time in Eq.~(\ref{time optimal}), and the qubit frequency in Eq.~(\ref{optimal omegaA}), one can obtain QFI of $\rho_P$ by Eq.~(\ref{QFI for mixed state}). Especially when the probe system is in a pure state, i.e., $c=|a_mb_m|$, we have $F_Q=4m^2(1-4a_m^2b_m^2\sin^2\phi_m)$, which is in agreement with Eq.~(\ref{QFI pure probe}). In case of a vanishing coherence, i.e., $c=0$, Eq.~(\ref{probeMixed}) becomes
\begin{equation}
\rho_P=a_m^2|j,m\rangle_{\rm opt}\langle j,m|+b_m^2|j,-m\rangle_{\rm opt}\langle j,-m|.
\end{equation}
Consequently, we have $F_Q=4m^2(1-4a_m^2b_m^2)$. It verifies that QFI can reach its maximum value $F_Q=N^2$ when the probe state is initialized as $|j,\pm j\rangle_{\rm opt}$ as indicated by Eq.~(\ref{state optimal}).

\begin{figure}[htbp]
\begin{centering}
\includegraphics[width=0.8\linewidth]{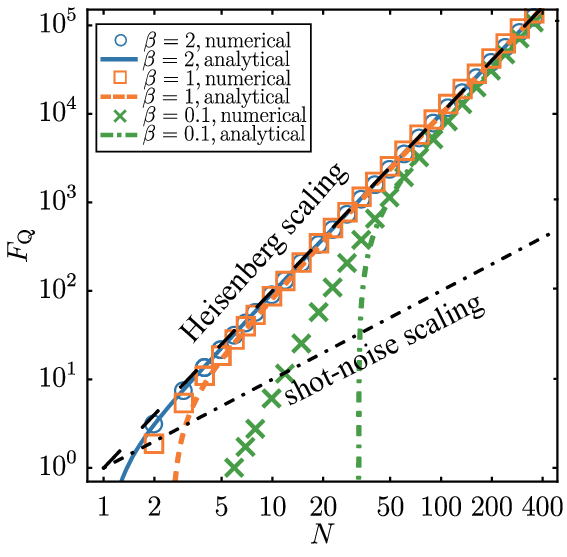}
\caption{QFI as a function of $N$ for a thermal state $\rho_P^{\rm th}$. The blue circles, orange squares, and green crosses indicate the exact numerical simulations in Eq.~(\ref{numerical thermal}) with inverse temperature $\beta=2$, $1$, and $0.1$, respectively. The blue solid line, orange dashed line, and green dot-dashed line are the corresponding approximate results given by Eq.~(\ref{QFI ana thermal}). The black-dashed line and the black dot-dashed line indicate the Heisenberg and shot-noise scalings, respectively. Here $\rho_A=|+\rangle\langle+|$, $t_1=t_{1,{\rm opt}}(n_1=0)=\pi/(2g)$, $\omega_P=10g$, and $\omega_A\approx5g$.}\label{QFI thermal state}
\end{centering}
\end{figure}

The probe system with no coherence can be extended to the case of a thermal state, i.e.,
\begin{equation}\label{thermal state}
\rho_P^{\rm th}=\sum_{m=-j}^j\frac{e^{-m\beta}}{Z}|j,m\rangle_{\rm opt}\langle j,m|,
\end{equation}
where $Z={\rm Tr}[\exp(-\beta J_{\rm opt})]$ is the partition function and $\beta\equiv\omega_P/(k_BT)$ is the dimensionless inverse temperature. Then with $d=N+1=2j+1$, $p_m=e^{-m\beta}/Z$, and $|\psi_m\rangle=|j,m\rangle_{\rm opt}$, Eq.~(\ref{QFI for mixed state}) becomes
\begin{equation}\label{numerical thermal}
\begin{aligned}
&F_Q=\frac{4}{Z}\sum_{m=-j}^jm^2e^{-m\beta}-\frac{8}{Z}\sum_{m=-j}^j\frac{m^2}{e^{-m\beta}+e^{m\beta}}\\
\geq&\frac{4\sum_{m=-j}^jm^2e^{-m\beta}}{\sum_{m=-j}^je^{-m\beta}}-\frac{\pi^3}{\beta^3\sum_{m=-j}^je^{-m\beta}}\\
=&\frac{1}{(e^\beta-1)^2}\Big[N^2\frac{1-e^{(3+N)\beta}}{1-e^{(1+N)\beta}}+(N+2)^2\frac{e^{2\beta}-e^{(1+N)\beta}}{1-e^{(1+N)\beta}}\\
+&(N^2+2N-2)\frac{2e^{(2+N)\beta}-2e^{\beta}}{1-e^{(1+N)\beta}}\Big]-\frac{\pi^3}{\beta^3}\frac{e^{\frac{N\beta}{2}}(e^\beta-1)}{e^{(N+1)\beta}-1},
\end{aligned}
\end{equation}
where we have used the inequality $\sum_{m=-j}^j\frac{m^2}{e^{-m\beta}+e^{m\beta}}\leq\sum_{m=-\infty}^\infty\frac{m^2}{e^{-m\beta}+e^{m\beta}}\approx\int_{-\infty}^\infty dm\frac{m^2}{e^{-m\beta}+e^{m\beta}}=\pi^3/(8\beta^3)$. Then for a large-size probe, i.e., $N\gg 1$, it is approximated as
\begin{equation}\label{QFI ana thermal}
F_Q\geq N^2-\frac{4}{e^\beta-1}N+\frac{4( e^\beta+1)}{(e^{\beta}-1)^2}-\frac{\pi^3}{\beta^3}e^{-\frac{N\beta}{2}}(1-e^{-\beta}).
\end{equation}
Clearly, $F_Q\rightarrow N^2$ when $\beta\rightarrow\infty$. It implies that the Heisenberg-scaling behavior dominates QFI at least in the low-temperature limit. The compact expression in Eq.~(\ref{QFI ana thermal}) is confirmed by the numerical result in Fig.~\ref{QFI thermal state} and they both approach the Heisenberg scaling for a sufficient large $N$. When $\beta=2$, $\beta=1$, and $\beta=0.1$, the approximated analytical results cannot be distinguished from the numerical simulation for $N\geq3$, $N\geq6$, and $N\geq52$, respectively. Even in a high-temperature case, i.e., $\beta=0.1$, QFI becomes following the square scaling law for $N>120$. With no loss of generality, we set $\omega_P=10g$ and $\omega_A\approx5g$ with $n_3=(9-N)/2$, $(10-N)/2$ for odd and even $N$, respectively, according to Eq.~(\ref{optimal omegaA}).

It should be emphasized that Eq.~(\ref{QFI for pure state}) presents the main advantage of our measurement-based metrology over the standard schemes~\cite{giovannetti2011advances,toth2014quantum,nawrocki2015introduction,polino2020photonic} described by Eq.~(\ref{QFI}). The unconditional measurement on the ancillary qubit establishes a special effective phase generator, i.e, $\mathcal{H}=e^{-i\pi J_z}J_{\rm opt}$. In the absence of the ancillary qubit, i.e., upon setting the coupling strength between the probe system and the ancillary qubit $g=0$, Eq.~(\ref{QFI for mixed state}) is rewritten as
\begin{equation}
F_Q=\sum_{i=1}^d4p_i\langle\psi_i|J_\phi^2|\psi_i\rangle
-\sum_{i,j=1}^d\frac{8p_ip_j}{p_i+p_j}|\langle\psi_i|J_\phi|\psi_j\rangle|^2,
\end{equation}
where the effective phase generator is $J_\phi\equiv\cos(\phi)J_x-\sin(\phi)J_y$ with $\phi=\omega_Pt_1$. In contrast to Eq.~(\ref{QFI for mixed state}), $F_Q=N^2$ is now attainable if and only if the probe is initialized as a GHZ state, i.e., $|\psi\rangle=(|j,j\rangle_\phi+e^{-i\phi_0}|j,-j\rangle_\phi)/\sqrt2$ with an arbitrary phase $\phi_0$. Here $|j,m\rangle_\phi$'s with $-j\leq m\leq j$ are the eigenstates of the collective angular momentum operator $J_\phi$.

Our protocol can still exceed the standard quantum limit of metrology in the simulation with experimental parameters for magnon-qubit systems~\cite{tabuchi2015coherent,lachance2020entanglement,xu2023quantum}. In the Appendix, one can see that under the dissipation environment, QFI gradually deviates from the Heisenberg scaling, drops below the standard quantum limit with increasing the probe spin number $N$, and eventually restores the linear scaling behavior as the probe system declines to the ground state.

\subsection{Effects of ancilla coherence}\label{QFI coherence qubit}

In the presence of a dephasing channel, it is reasonable to assume that the ancillary qubit becomes
\begin{equation}\label{ancillary mixed}
\rho_A=c_+^2|+\rangle\langle+|+c_-^2|-\rangle\langle-|+c(e^{-i\phi_A}|+\rangle\langle-|+{\rm H.c.}),
\end{equation}
where $c_+$, $c_-$, and $c$ are positive numbers with $c_+^2+c_-^2=1$ and $0\leq c\leq |c_+c_-|$, and $\phi_A$ is an arbitrary phase. The probe system is supposed to be prepared as $|\psi\rangle=|j,j\rangle_{\rm opt}$ and the joint-evolution time during Evolution 1 is optimized as $t_1=t_{1,{\rm opt}}(n_1=0)=\pi/(2g)$.

In case of a finite coherence $c>0$, if the measurement result about the ancillary qubit is $|+\rangle$, then the output state under the whole evolution operator in Eq.~(\ref{U}) will be
\begin{equation}\label{outputState}
\begin{aligned}
\rho_{\theta,+}&=\mathcal{N}^{-1}_+U_{\theta,+}|j,j\rangle_{\rm opt}\langle j,j|\otimes\rho_AU_{\theta,+}^\dagger\\
&=[c_+^2|\Psi_{\theta,+}\rangle\langle\Psi_{\theta,+}|+c_-^2|\Psi_{\theta,-}\rangle\langle\Psi_{\theta,-}|\\
&+c(e^{-i\phi_A}|\Psi_{\theta,+}\rangle\langle\Psi_{\theta,-}|+{\rm H.c.})]\otimes|+\rangle\langle+|
\end{aligned}
\end{equation}
with a probability $\mathcal{N}_+=1/2$, where the normalized probe states read
\begin{equation}
|\Psi_{\theta,\pm}\rangle=\frac{1}{\sqrt{2}}U(t_2)R_x(\theta)
\left(|j,j\rangle_x\pm i^Ne^{i\frac{\pi}{g}\omega_A}|j,-j\rangle_x\right).
\end{equation}
On diagonalization, the output state in Eq.~(\ref{outputState}) becomes
\begin{equation}
\rho_{\theta,+}=c_1^2|\Psi_1\rangle\langle\Psi_1|+c_2^2|\Psi_2\rangle\langle\Psi_2|,
\end{equation}
where $c_{1/2}^2=(1\pm\Omega)/2$ with $\Omega=\sqrt{4c^2+\Delta^2}$ and $\Delta=c_+^2-c_-^2$ and
\begin{equation}
|\Psi_{1/2}\rangle=\frac{(\Delta\pm\Omega)
e^{-i\phi_A}|\Psi_{\theta,+}\rangle+2c|\Psi_{\theta,-}\rangle}{\sqrt{4c^2+(\Delta\pm\Omega)^2}}\otimes|+\rangle.
\end{equation}
Using the mixed-state formula for QFI in Eq.~(\ref{QFI after encoding}), one can obtain the effective quantum Fisher information $F_{Q,+,{\rm eff}}$. Similarly, when the unconditional measurement on the ancillary qubit yields $|-\rangle$, we can have $F_{Q,-,{\rm eff}}$. When the qubit is a pure state, i.e., $c=|c_+c_-|$, we have $c_1^2=1$, $c_2^2=0$, and the full QFI becomes
\begin{equation}
\begin{aligned}
F_Q&=F_{Q,+,{\rm eff}}+F_{Q,-,{\rm eff}} \\
&=c_1^2F_{Q,1}+c_2^2F_{Q,2}-16c_1^2c_2^2|\langle\Psi_1|\partial_\theta\Psi_2\rangle|^2\\
&=N^2(1-4c_+^2c_-^2\sin^2\phi_A).
\end{aligned}
\end{equation}
It is found that the Heisenberg limit of QFI can be reached when $c_+c_-=0$, that is equivalent to Eq.~(\ref{ancillary optimal}), or when $\sin\phi_A=0$. The latter identifies more optimized states of the ancillary system for probe metrology with the exact Heisenberg limit, i.e., the ancillary qubit can be prepared in a full coherent state with $\phi_A=n_4\pi$ and an integer $n_4$.

In case of zero coherence, i.e., $c=0$, the output state in Eq.~(\ref{outputState}) on the measurement result $|+\rangle$ becomes
\begin{equation}
\rho_{\theta,+}=(c_+^2|\Psi_{\theta,+}\rangle\langle\Psi_{\theta,+}|+c_-^2|\Psi_{\theta,-}\rangle\langle\Psi_{\theta,-}|)\otimes|+\rangle\langle+|,
\end{equation}
with a probability $\mathcal{N}_+=1/2$. Using Eq.~(\ref{QFI after encoding}), the effective QFI becomes
\begin{equation}
F_{Q,+,{\rm eff}}=\frac{N^2}{2}\left(1-4c_1^2c_2^2\right).
\end{equation}
When the unconditional measurement on the ancillary qubit yields $|-\rangle$, one can obtain the same result. Hence the full QFI is $F_Q=F_{Q,+,{\rm eff}}+F_{Q,-,{\rm eff}}=N^2(1-4c_+^2c_-^2)$. Thus QFI still follows the Heisenberg scaling $F_Q\propto N^2$ even if the qubit is prepared as a completely mixed state with $c_+c_-\neq1/2$, which otherwise describes the useless infinite high-temperature state.

\section{quantum Fisher information under imprecise control}\label{ImperfectControl}

The precise control over the joint evolution time $t_1$ in Eq.~(\ref{time optimal}) and the exact synchronization of the parametric encoding and the unconditional measurement on the ancillary qubit described by the whole unitary evolution in Eq.~(\ref{U}) are not strictly required in our protocol.

\begin{figure}[htbp]
\begin{centering}
\includegraphics[width=0.9\linewidth]{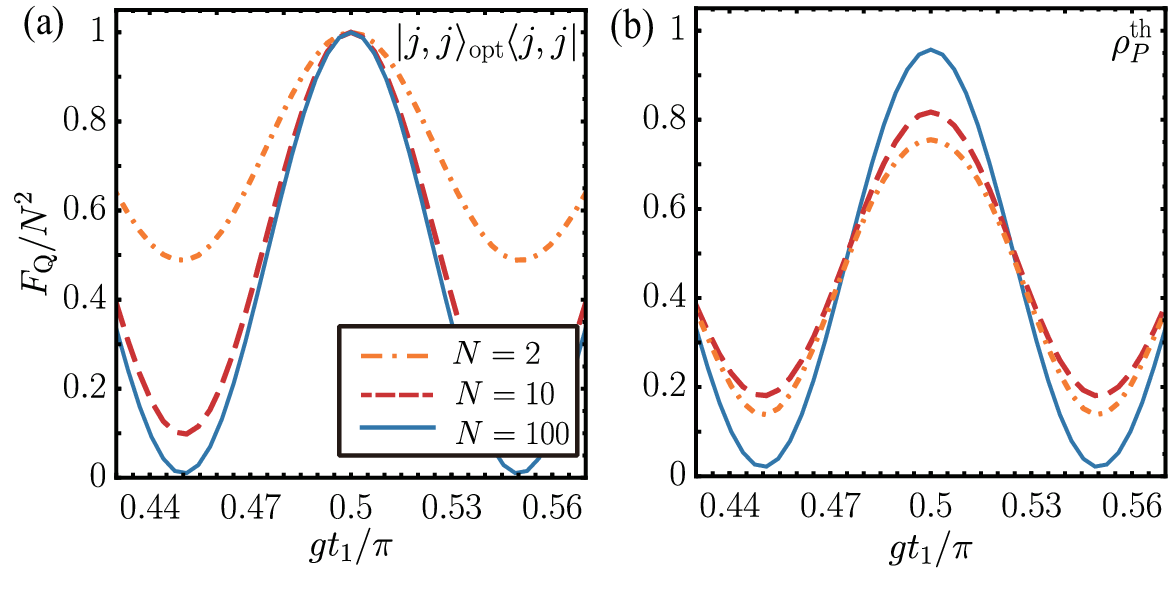}
\caption{Renormalized QFI $F_Q/N^2$ as a function of arbitrary periods $gt_1$ under various probe-spin number $N$ for the probe state prepared as (a) the polarized state $\rho_P=|j,j\rangle_{\rm opt}\langle j,j|$ or (b) the thermal state $\rho^{\rm th}_P$ with $\beta=1$. The other parameters are the same as Fig.~\ref{QFI thermal state}.}\label{QFI evolution time}
\end{centering}
\end{figure}

With the ancillary-qubit state in Eq.~(\ref{ancillary optimal}) and the probe state in Eq.~(\ref{state optimal}), the unnormalized output state in Eq.~(\ref{output state}) becomes
\begin{equation}
|\Psi_{\theta,\pm}\rangle=e^{-iHt_2}e^{-i\theta J_x}|\pm\rangle\langle\pm|e^{-iHt_1}|j,j\rangle_{\rm opt}\otimes|+\rangle
\end{equation}
under an arbitrary $t_1$, which can be normalized by the probability $\mathcal N_\pm=[1\pm\cos(2\omega_At_1)\cos^N(gt_1)]/2$. Through a similar calculation by Eq.~(\ref{QFI pm}), one can find that $F_Q=N^2$ if and only if $t_1=t_{1,{\rm opt}}$ given in Eq.~(\ref{time optimal}).

This result can be confirmed by the numerical simulation over the renormalized QFI as a function of the joint evolution time $t_1$ as shown in Figs.~\ref{QFI evolution time}(a) and \ref{QFI evolution time}(b), where the probe (spin ensemble) is initialized as the polarized state $\rho_P=|j,j\rangle_{\rm opt}\langle j,j|$ and the thermal state $\rho_P^{\rm th}$ with $\beta=1$, respectively. It is found that QFI presents a central-symmetrical pattern around the optimized point $gt_1=\pi/2$. For the polarized state, $F_Q$ reaches the peak value $N^2$ when $gt_1=\pi/2$, irrespective of the probe size $N$. For the thermal state, $F_Q$ is proportional to $N^2$ when $gt_1=\pi/2$ and the scale factor increases with $N$. In particular, we have $F_Q/N^2\approx0.75$, $0.82$, and $0.96$ for $N=2$, $10$, and $100$, respectively.

In both Figs.~\ref{QFI evolution time}(a) and \ref{QFI evolution time}(b), $F_Q$ varies smoothly around the optimized point, indicating that our metrology protocol is not sensitive to the imprecise control over the joint evolution period $t_1$ or the coupling strength $g$. For a large-size probe, i.e., $N\gg1$, when $|\cos(gt_1)|\neq1$, we have
\begin{equation}\label{QFI approximation}
\begin{aligned}
F_{Q,\pm,{\rm eff}}&\approx\frac{N^2}{4}\left\{1-\cos(2gt_1)\cos\left[2\omega_P\left(\frac{\pi}{2g}-t_1\right)\right]\right.\\
&\left.-2\left|\cos(gt_1)\sin\left[\omega_P\left(\frac{\pi}{2g}-t_1\right)\right]\right|^2\right\}
\end{aligned}
\end{equation}
according to Eq.~(\ref{QFI pm}). Substituting $t_1=t_{1,{\rm opt}}(n_1=0)+\delta t_1=\pi/(2g)+\delta t_1$ with $|\delta t_1/t_{1,{\rm opt}}(n_1=0)|\ll1$, the full QFI in Eq.~(\ref{QFI optimal}) can be written as
\begin{equation}\label{sensitivity time}
F_Q=F_{Q,+,{\rm eff}}+F_{Q,-,{\rm eff}}\approx N^2[1-(g^2+\omega_P^2)\delta t_1^2]
\end{equation}
up to the second order in the timing deviation $\delta t_1$. In addition, when $t_1=t_{1,{\rm opt}}(n_1)$ and $g\rightarrow g+\Delta g$ with $|\Delta g/g|\ll1$, QFI can be written as
\begin{equation}
F_Q\approx N^2\left[1-\left(1+\frac{4\omega_P^2t_{1,{\rm opt}}^2}{\pi^2}\right)\Delta g^2t_{1,{\rm opt}}^2\right]
\end{equation}
up to the second order in the deviation $\Delta g$.

\begin{figure}[htbp]
\begin{centering}
\includegraphics[width=0.9\linewidth]{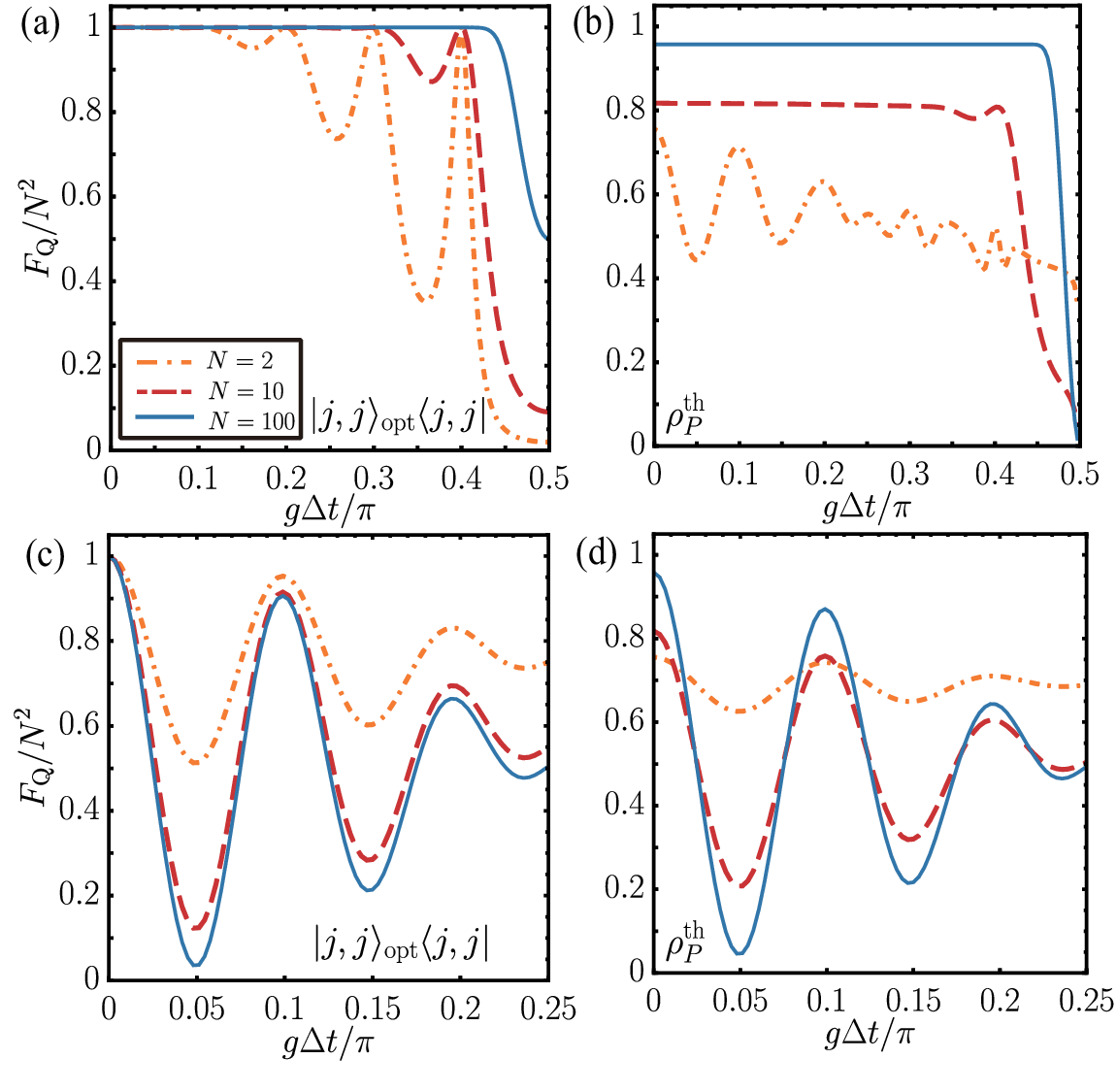}
\caption{(a), and (b) Renormalized QFI $F_Q/N^2$ as a function of the measurement delay $g\Delta t$ under various probe-spin numbers $N$. (c), and (d) $F_Q/N^2$ as a function of the parametric encoding delay $g\Delta t$ under various $N$. In (a) and (c), the probe state is prepared as a polarized state $\rho_P=|j,j\rangle_{\rm opt}\langle j,j|$; in (b) and (d), it starts from a thermal state $\rho^{\rm th}_P$ with $\beta=1$. $gt_1=\pi/2$ and the other parameters are the same as Fig.~\ref{QFI thermal state}.}\label{QFI delay}
\end{centering}
\end{figure}

The synchronization of encoding and measurement is broken when a time delay occurs between them. If the measurement falls behind the encoding with an interval $\Delta t$, the whole evolution operator of the circuit in Eq.~(\ref{U}) is then modified to be
\begin{equation}
\begin{aligned}
U_{\theta,\pm}&=U(t_2)M_\pm U(\Delta t)R_x(\theta)U(t_{1,{\rm opt}})\\
& =e^{-iHt_2}|\pm\rangle\langle\pm|e^{-iH\Delta t}e^{-i\theta J_x}e^{-iHt_{1,{\rm opt}}},
\end{aligned}
\end{equation}
with the optimized $t_1$ in Eq.~(\ref{time optimal}). Consequently, the output state in Eq.~(\ref{output state}) becomes
\begin{equation}
\begin{aligned}
&|\Psi_{\theta,\pm}\rangle=U_{\theta,\pm}|j,j\rangle_{\rm opt}\otimes|+\rangle\\
=&\frac{U(t_2)}{2}e^{-i(\omega_P+g)J_z\Delta t}[e^{-i\omega_A(\Delta t+t_{1,{\rm opt}})}e^{-ij\theta}|j,j\rangle_x\\
&\pm i^Ne^{2igJ_z\Delta t}e^{i\omega_A(\Delta t+t_{1,{\rm opt}})}e^{ij\theta}|j,-j\rangle_x]\otimes|\pm\rangle,
\end{aligned}
\end{equation}
upon the initial qubit state $|\varphi\rangle_{\rm opt}=|+\rangle$. As an analog to Eqs.~(\ref{QFI pm}) and (\ref{QFI optimal}), we have
\begin{equation}\label{QFI measurement delay}
\begin{aligned}
&F_Q=F_{Q,+,{\rm eff}}+F_{Q,-,{\rm eff}}\\
=&\frac{N^2\left[1-\sin ^{2N}(g\Delta t)\right]}{1-\sin^{2N}(g\Delta t)\cos^2\left[2\omega_A(\Delta t+t_{1,{\rm opt}})+N\theta\right]}.
\end{aligned}
\end{equation}
This expression indicates that QFI turns out to be dependent on the to-be-estimated phase $\theta$ under general situations. For a large-size probe, i.e., $N\gg1$, the dependence of QFI on $\theta$ gradually vanishes and Eq.~(\ref{QFI measurement delay}) can be approximated as $F_Q\approx N^2$ when $|\sin(g\Delta t)|<1$. Without loss of generality, we set $\theta\approx0$ and show the dependence of QFI on the time delay $\Delta t$ of measurement under various probe size $N$ in Fig.~\ref{QFI delay}(a) for the polarized state of probe. One can find that the quantum Fisher information tends to maintain its maximum value $N^2$ in a wider regime of $\Delta t$ for a larger probe size $N$, as illustrated by Eq.~(\ref{QFI measurement delay}). This reduction still holds when the probe starts from a thermal state, as shown in Fig.~\ref{QFI delay}(b). It indicates that a large probe can significantly reduce its sensitivity to the imprecise control over the measurement moment.

As for the nonvanishing encoding delay, i.e., the parametric encoding falls behind the measurement with an interval $\Delta t$, the whole evolution operator in Eq.~(\ref{U}) becomes
\begin{equation}
\begin{aligned}
U_{\theta,\pm}&=U(t_2)R_x(\theta)U(\Delta t)M_\pm U(t_{1,{\rm opt}})\\
& =e^{-iHt_2}e^{-i\theta J_x}e^{-iH\Delta t}|\pm\rangle\langle\pm|e^{-iHt_{1,{\rm opt}}},
\end{aligned}
\end{equation}
when the other conditions are invariant.

Upon the optimal input states in Eqs.~(\ref{ancillary optimal}) and (\ref{state optimal}), Eq.~(\ref{output state}) is now rewritten as
\begin{equation}
\begin{aligned}
|\Psi_{\theta,\pm}\rangle&=U_{\theta,\pm}|j,j\rangle_{\rm opt}\otimes|+\rangle\\
&=\frac{U(t_2)}{2}R_x(\theta)U(\Delta t)e^{-i\omega_At_{1,{\rm opt}}}\\
&\times(\mathcal I^{N+1}\pm e^{2i(\omega_A+gJ_z)t_{1,{\rm opt}}})|j,j\rangle_x\otimes|\pm\rangle,
\end{aligned}
\end{equation}
with a fixed probability $\mathcal N_{\theta,\pm}=1/2$. Consequently, the effective QFI is
\begin{equation}
\begin{aligned}
F_Q&=F_{Q,+,{\rm eff}}+F_{Q,-,{\rm eff}}\\
&=\frac{N}{4}(2(N+1)+(N-1)\\
&\times\{\cos[2\Delta t(g-\omega_P)]+\cos[2\Delta t(g+\omega_P)]\}).
\end{aligned}
\end{equation}
It is consistent with Eq.~(\ref{QFI optimal}) that $F_Q=N^2$ when $\Delta t=0$. When $\Delta t\ll1$, we have
\begin{equation}
F_Q\approx N^2-N(N-1)(g^2+\omega_P^2)\Delta t^2,
\end{equation}
up to the second order of $\Delta t$. Again it confirms that our condition about the joint evolution time $t_1$ in Eq.~(\ref{time optimal}) is optimal. In Figs.~\ref{QFI delay}(c) and \ref{QFI delay}(d), we show the dependence of $F_Q$ on the encoding delay $\Delta t$ for the polarized probe state and the thermal probe state, respectively, under the ancillary qubit state in Eq.~(\ref{ancillary optimal}) and the optimal evolution time $t_1$ in Eq.~(\ref{time optimal}). Both of them demonstrate that the detrimental effect from the encoding decay on QFI fluctuates with the time delay of encoding.

\section{Classical Fisher information}\label{ClassicalFisherInf}

In a practical parametric estimation, CFI measures the amount of information that can be extracted from the probability distribution of the output state~\cite{braun2018quantum,liu2020quantum,tan2021fisher}, which is upper bounded by QFI about the metrology precision. In this section, we derive CFI of our protocol under the same settings as those for QFI in Fig.~\ref{QFI thermal state}, i.e., the optimized joint-evolution time $t_{1,{\rm opt}}$, and an optimal state of the probe system $|j,j\rangle_{\rm opt}$, and the ancillary qubit $|\varphi\rangle_{\rm opt}$ in Eq.~(\ref{time optimal}) with $n_1=0$, Eq.~(\ref{state optimal}), and Eq.~(\ref{ancillary optimal}), respectively.

\begin{figure}[htbp]
\begin{centering}
\includegraphics[width=0.9\linewidth]{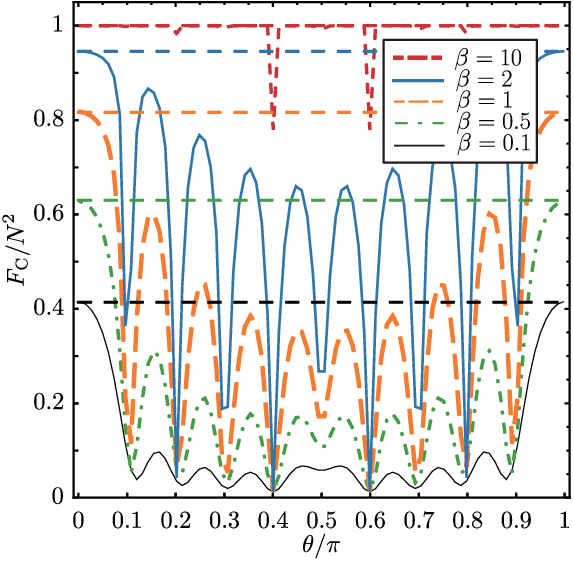}
\caption{Renormalized CFI $F_C/N^2$ as a function of the to-be-estimated phase $\theta$ for the probe state prepared as a thermal state $\rho^{\rm th}_P$ with various $\beta$. The dashed horizontal lines indicate the corresponding renormalized QFI $F_Q/N^2$. The ancillary qubit is initialized as $\rho_A=|+\rangle\langle +|$. The probe size is $N=10$. The other parameters are the same as Fig.~\ref{QFI thermal state}.}\label{CFI thermal}
\end{centering}
\end{figure}

The probe state after the three-stage evolution as described by the evolution operator~(\ref{U}) reads
\begin{equation}
\begin{aligned}
\rho_\pm(\theta)=&{\rm Tr}_A[U_{\theta,\pm}\rho_P\otimes|+\rangle\langle+|U_{\theta,\pm}^\dagger]\\
=&\frac{1}{2}{\rm Tr}_A\{U(t_2)[|j,j\rangle_x\langle j,j|+|j,-j\rangle_x\langle j,-j|\\
&\pm e^{-i\Phi}|j,j\rangle_x\langle j,-j|\pm e^{i\Phi}|j,-j\rangle_x\langle j,j|]\\
&\otimes|\pm\rangle\langle\pm|U^\dagger(t_2)\},
\end{aligned}
\end{equation}
with the normalization coefficient $\mathcal N_{\theta,\pm}=1/2$ and the phase $\Phi\equiv 2\omega_At_{1,{\rm opt}}+N\theta+N\pi/2$. Subsequently, one can perform the projective measurements $|j,m\rangle\langle j,m|$ on the probe system as the last operation of the circuit model in Fig.~\ref{protocol}, where $|j,m\rangle$ denotes the eigenstate of the collective spin operators $J_z$ with eigenvalue $m$. The probability of detecting the probe state in the state $|j,m\rangle$ and its first derivative with respect to the to-be-estimated phase $\theta$ read
\begin{equation}
\begin{aligned}
&P(m,\pm|\theta)=\langle j,m|\rho_\pm(\theta)|j,m\rangle\\
=&\frac{1}{2}\left[|\langle j,m|j,j\rangle_x|^2\pm(-i)^{2(j+m)}\cos\Phi|\langle j,m|j,j\rangle_x|^2\right],\\
&\partial_\theta P(m,\pm|\theta)=\mp\frac{(-i)^{2(j+m)}N}{2}\sin\Phi|\langle j,m|j,j\rangle_x|^2,
\end{aligned}
\end{equation}
respectively. Consequently, we have
\begin{equation}\label{partial derivative}
\begin{aligned}
&\frac{[\partial_\theta P(m,\pm|\theta)]^2}{P(m,\pm|\theta)}=\frac{N^2}{2}\frac{\sin^2\Phi|\langle j,m|j,j\rangle_x|^2}{1\pm(-i)^{2(j+m)}\cos\Phi}\\
=&\frac{N^2}{2}[1\mp i^{2(j-m)}\cos\Phi]|\langle j,m|j,j\rangle_x|^2.
\end{aligned}
\end{equation}
In the proceeding two steps, we have used the identities
\begin{equation}
  \begin{aligned}
   &\langle j,m|j,-j\rangle_x=(-i)^{2(j+m)}\langle j,m|j,j\rangle_x, \\
   & \sin^2\Phi=[1\pm(-i)^{2(j+m)}\cos\Phi][1\mp i^{2(j-m)}\cos\Phi].
  \end{aligned}
\end{equation}
Substituting Eq.~(\ref{partial derivative}) to Eq.~(\ref{CFI}), we have
\begin{equation}
\begin{aligned}
F_C&=\sum_{m=-j}^j\left(\frac{[\partial_\theta P( m,+|\theta)]^2}{P(m,+|\theta)}+\frac{[\partial_{\theta}P(m,-|\theta)]^2}{P(m,-|\theta)}\right)\\
&=N^2\sum_{m=-j}^j|\langle j,m|j,j\rangle_x|^2=N^2=F_Q.
\end{aligned}
\end{equation}
The last line is exactly the same as Eq.~(\ref{QFI optimal}), irrespective of the idle evolution time $t_2$. The derivation also applies to $|j,-j\rangle_{\rm opt}$ and $a|j,j\rangle_{\rm opt}+be^{-i\phi}|j,-j\rangle_{\rm opt}$. It is thus verified that in our protocol CFI can saturate with its quantum counterpart as long as the probe is prepared as an optimized state.

In Fig.~\ref{CFI thermal}, we plot the renormalized CFI $F_C/N^2$ as a function of the to-be-estimated parameter $\theta$ for the thermal state $\rho^{\rm th}_P$ with a fixed probe spin number $N=10$ and various $\beta$. It is found that, in the high-temperature limit, i.e., $\beta\rightarrow0$, the classical Fisher information is much lower than the quantum Fisher information (see the lowest horizontal line for $\beta=0.1$), except when $\theta$ is around $0$ and $\pi$. On the contrary, in the low-temperature limit (see the highest horizontal line), e.g., $\beta=10$, CFI saturates with its quantum counterpart across almost the whole regime of $\theta$. In addition, the upperbound of CFI is found to be $F_C/N^2=0.415$, $0.628$, $0.818$, $0.946$, and almost unit for $\beta=0.1$, $0.5$, $1$, $2$, and $10$, respectively.

\section{conclusion}\label{conclusion}

In summary, we introduce the measurement on an ancillary system as an unconventional resource to replace entanglement or nonlinear Hamiltonian in standard quantum metrology protocols performing in large-spin or spin-ensemble systems. Within our metrology protocol by measurement, the polarized states of an optimized collective angular momentum operator and their superposition can be used to exactly attain the Heisenberg-scaling in parameter estimation with respect to the probe size $N$. Our protocol is of the strategies by the probe-ancilla interaction, that allows the probe units and ancillary system initially prepared as the thermal or mixed state to have an asymptotic square scaling law of metrology if $N$ is sufficiently large. The metrology precision by our protocol can be optimized under the precise control over the joint evolution time of probe and ancilla before the exact synchronization of the parametric encoding on the probe and the measurement on the qubit. The numerical simulation shows that it is not sensitive to the fluctuation of the joint-evolution time and the time delay between encoding and measurement. By virtue of the projective measurement on the probe system, the classical Fisher information in our protocol can saturate with its quantum counterpart, irrespective of the idle evolution time after the parametric encoding. In essence, the developed protocol confirms that both GHZ-like state and nonlinear Hamiltonian are not necessary conditions to achieve the Heisenberg-scaling metrology. We can have an economical way, e.g., using an almost classical state, to achieve the precision exceeding standard quantum limit.

\section*{Acknowledgments}

We acknowledge grant support from the Science and Technology Program of Zhejiang Province (No. 2025C01028).

\section*{Appendix: QFI under dissipation}\label{appendix dissipative environment}

\begin{figure}[htbp]
\begin{centering}
\includegraphics[width=0.8\linewidth]{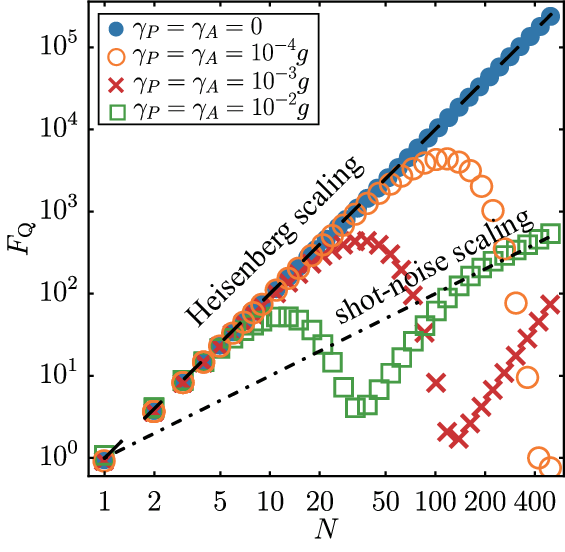}
\caption{QFI as a function of $N$ for a thermal state $\rho_P^{\rm th}$ under a dissipative environment with various probe and qubit decay rates. $\omega_P=\omega_A=10^3g$, $\beta=10$, and the other parameters are the same as Fig.~\ref{QFI thermal state}.}\label{QFIRealParameters}
\end{centering}
\end{figure}

This appendix is devoted to the implementation of our metrology protocol with real experimental parameters of the hybrid magnon-qubit systems~\cite{tabuchi2015coherent,lachance2020entanglement,xu2023quantum}, where the magnon and the superconducting qubit play the role of the probe spin ensemble and the ancillary system, respectively. The interaction Hamiltonian in Eq.~(\ref{H}) stems from a dispersive interaction between the Kittel mode of the magnon and the qubit,
\begin{equation}\label{dispersiveInteraction}
H_I=g\sigma_zm^\dagger m=Ng\sigma_z-gJ_z\sigma_z,
\end{equation}
where we applied HP transformation~\cite{kounalakis2022analog,rameshti2022cavity}, $m^\dagger m=N-J_z$. $m$ and $m^\dagger$ are the annihilation and creation operators of the magnon, respectively, $J_z$ is the collective spin operator, and $N$ denotes the magnon spin number. Substituting Eq.~(\ref{dispersiveInteraction}) into Eq.~(\ref{H}), the full Hamiltonian becomes
\begin{equation}
H=H_0+H_I=\omega_pJ_z+\omega'_A\sigma_z-gJ_z\sigma_z,
\end{equation}
where the effective qubit frequency denotes $\omega'_A=\omega_A+Ng$. In the presence of a dissipative environment close to zero temperature, the evolution of the density matrix $\rho(t)$ of the composite system can be described by the master equation,
\begin{equation}\label{masterEquation}
\dot\rho(t)=-i[H, \rho(t)]+\gamma_P\mathcal L[J_-]\rho(t)+\gamma_A\mathcal L[\sigma_-]\rho(t),
\end{equation}
where $\gamma_P$ and $\gamma_A$ are the probe and qubit decay rates, respectively, and the Lindblad superoperator $\mathcal L$ is defined as $\mathcal L[O]\rho\equiv\frac{1}{2}(2O\rho O^\dagger-O^\dagger O\rho-\rho O^\dagger O)$ with $O=J_-$ or $\sigma_-$.

In Fig.~\ref{QFIRealParameters}, we present the quantum Fisher information under a dissipative environment with various decay rates, where the probe and qubit are initialized as the thermal state and pure state in Eqs.~(\ref{thermal state}) and (\ref{ancillary optimal}), respectively. According to the experimental parameters in magnon-qubit systems~\cite{tabuchi2015coherent,lachance2020entanglement,xu2023quantum}, it is reasonable to set $\omega_P\approx \omega_A\approx10^3g$ and $\beta\approx10$ in numerical simulation. In the absence of dissipation, i.e., $\gamma_P=\gamma_A=0$, it is found that QFI can exactly follow the square scaling $F_Q=N^2$, which is irrelevant to the ratio of the coupling strength and system frequency and is in agreement with Eq.~(\ref{QFI ana thermal}). In the presence of the dissipative, the Heisenberg scaling still dominates for $N\leq10$, $N\leq22$, and $N\leq71$ under $\gamma_P/g=\gamma_A/g=10^{-2}$, $10^{-3}$, and $10^{-4}$, respectively. With increasing probe spin number $N$, the scaling behavior of QFI gradually deviates from the Heisenberg scaling and drops to the shot-noise scaling. In particular, we have $F_Q\approx N$ for $N=21$, $N=78$, and $N=271$, for $\gamma_P/g=\gamma_A/g=10^{-2}$, $10^{-3}$, and $10^{-4}$, respectively. Despite $F_Q$ keeps declining for a short range of $N$, it will eventually restore the standard quantum limit, since the probe system approaches the ground state $|j,-j\rangle$ under dissipation.

\bibliographystyle{apsrevlong}
\bibliography{ref}

\begin{thebibliography}{65}%
\makeatletter
\providecommand \@ifxundefined [1]{%
 \@ifx{#1\undefined}
}%
\providecommand \@ifnum [1]{%
 \ifnum #1\expandafter \@firstoftwo
 \else \expandafter \@secondoftwo
 \fi
}%
\providecommand \@ifx [1]{%
 \ifx #1\expandafter \@firstoftwo
 \else \expandafter \@secondoftwo
 \fi
}%
\providecommand \natexlab [1]{#1}%
\providecommand \enquote  [1]{``#1''}%
\providecommand \bibnamefont  [1]{#1}%
\providecommand \bibfnamefont [1]{#1}%
\providecommand \citenamefont [1]{#1}%
\providecommand \href@noop [0]{\@secondoftwo}%
\providecommand \href [0]{\begingroup \@sanitize@url \@href}%
\providecommand \@href[1]{\@@startlink{#1}\@@href}%
\providecommand \@@href[1]{\endgroup#1\@@endlink}%
\providecommand \@sanitize@url [0]{\catcode `\\12\catcode `\$12\catcode
  `\&12\catcode `\#12\catcode `\^12\catcode `\_12\catcode `\%12\relax}%
\providecommand \@@startlink[1]{}%
\providecommand \@@endlink[0]{}%
\providecommand \url  [0]{\begingroup\@sanitize@url \@url }%
\providecommand \@url [1]{\endgroup\@href {#1}{\urlprefix }}%
\providecommand \urlprefix  [0]{URL }%
\providecommand \Eprint [0]{\href }%
\providecommand \doibase [0]{http://dx.doi.org/}%
\providecommand \selectlanguage [0]{\@gobble}%
\providecommand \bibinfo  [0]{\@secondoftwo}%
\providecommand \bibfield  [0]{\@secondoftwo}%
\providecommand \translation [1]{[#1]}%
\providecommand \BibitemOpen [0]{}%
\providecommand \bibitemStop [0]{}%
\providecommand \bibitemNoStop [0]{.\EOS\space}%
\providecommand \EOS [0]{\spacefactor3000\relax}%
\providecommand \BibitemShut  [1]{\csname bibitem#1\endcsname}%
\let\auto@bib@innerbib\@empty
%</preamble>
\bibitem [{\citenamefont {Sun}\ \emph {et~al.}(2010)\citenamefont {Sun},
  \citenamefont {Ma}, \citenamefont {Lu},\ and\ \citenamefont
  {Wang}}]{Sun2010fisher}%
  \BibitemOpen
  \bibfield  {author} {\bibinfo {author} {\bibfnamefont {Z.}~\bibnamefont
  {Sun}}, \bibinfo {author} {\bibfnamefont {J.}~\bibnamefont {Ma}}, \bibinfo
  {author} {\bibfnamefont {X.-M.}\ \bibnamefont {Lu}}, \ and\ \bibinfo {author}
  {\bibfnamefont {X.-G.}\ \bibnamefont {Wang}},\ }\bibfield  {title} {\emph
  {\bibinfo {title} {Fisher information in a quantum-critical environment},\
  }}\href {\doibase 10.1103/PhysRevA.82.022306} {\bibfield  {journal} {\bibinfo
   {journal} {Phys. Rev. A}\ }\textbf {\bibinfo {volume} {82}},\ \bibinfo
  {pages} {022306} (\bibinfo {year} {2010})}\BibitemShut {NoStop}%
\bibitem [{\citenamefont {Ma}\ \emph {et~al.}(2011)\citenamefont {Ma},
  \citenamefont {Huang}, \citenamefont {Wang},\ and\ \citenamefont
  {Sun}}]{ma2011quantum}%
  \BibitemOpen
  \bibfield  {author} {\bibinfo {author} {\bibfnamefont {J.}~\bibnamefont
  {Ma}}, \bibinfo {author} {\bibfnamefont {Y.-X.}\ \bibnamefont {Huang}},
  \bibinfo {author} {\bibfnamefont {X.-G.}\ \bibnamefont {Wang}}, \ and\
  \bibinfo {author} {\bibfnamefont {C.~P.}\ \bibnamefont {Sun}},\ }\bibfield
  {title} {\emph {\bibinfo {title} {Quantum {F}isher information of the
  {Greenberger-Horne-Zeilinger} state in decoherence channels},\ }}\href
  {\doibase 10.1103/PhysRevA.84.022302} {\bibfield  {journal} {\bibinfo
  {journal} {Phys. Rev. A}\ }\textbf {\bibinfo {volume} {84}},\ \bibinfo
  {pages} {022302} (\bibinfo {year} {2011})}\BibitemShut {NoStop}%
\bibitem [{\citenamefont {Genoni}\ \emph {et~al.}(2012)\citenamefont {Genoni},
  \citenamefont {Olivares}, \citenamefont {Brivio}, \citenamefont {Cialdi},
  \citenamefont {Cipriani}, \citenamefont {Santamato}, \citenamefont
  {Vezzoli},\ and\ \citenamefont {Paris}}]{genoni2012optical}%
  \BibitemOpen
  \bibfield  {author} {\bibinfo {author} {\bibfnamefont {M.~G.}\ \bibnamefont
  {Genoni}}, \bibinfo {author} {\bibfnamefont {S.}~\bibnamefont {Olivares}},
  \bibinfo {author} {\bibfnamefont {D.}~\bibnamefont {Brivio}}, \bibinfo
  {author} {\bibfnamefont {S.}~\bibnamefont {Cialdi}}, \bibinfo {author}
  {\bibfnamefont {D.}~\bibnamefont {Cipriani}}, \bibinfo {author}
  {\bibfnamefont {A.}~\bibnamefont {Santamato}}, \bibinfo {author}
  {\bibfnamefont {S.}~\bibnamefont {Vezzoli}}, \ and\ \bibinfo {author}
  {\bibfnamefont {M.~G.~A.}\ \bibnamefont {Paris}},\ }\bibfield  {title} {\emph
  {\bibinfo {title} {Optical interferometry in the presence of large phase
  diffusion},\ }}\href {\doibase 10.1103/PhysRevA.85.043817} {\bibfield
  {journal} {\bibinfo  {journal} {Phys. Rev. A}\ }\textbf {\bibinfo {volume}
  {85}},\ \bibinfo {pages} {043817} (\bibinfo {year} {2012})}\BibitemShut
  {NoStop}%
\bibitem [{\citenamefont {Escher}\ \emph {et~al.}(2012)\citenamefont {Escher},
  \citenamefont {Davidovich}, \citenamefont {Zagury},\ and\ \citenamefont
  {de~Matos~Filho}}]{escher2012quantum}%
  \BibitemOpen
  \bibfield  {author} {\bibinfo {author} {\bibfnamefont {B.~M.}\ \bibnamefont
  {Escher}}, \bibinfo {author} {\bibfnamefont {L.}~\bibnamefont {Davidovich}},
  \bibinfo {author} {\bibfnamefont {N.}~\bibnamefont {Zagury}}, \ and\ \bibinfo
  {author} {\bibfnamefont {R.~L.}\ \bibnamefont {de~Matos~Filho}},\ }\bibfield
  {title} {\emph {\bibinfo {title} {Quantum metrological limits via a
  variational approach},\ }}\href {\doibase 10.1103/PhysRevLett.109.190404}
  {\bibfield  {journal} {\bibinfo  {journal} {Phys. Rev. Lett.}\ }\textbf
  {\bibinfo {volume} {109}},\ \bibinfo {pages} {190404} (\bibinfo {year}
  {2012})}\BibitemShut {NoStop}%
\bibitem [{\citenamefont {Zhong}\ \emph {et~al.}(2013)\citenamefont {Zhong},
  \citenamefont {Sun}, \citenamefont {Ma}, \citenamefont {Wang},\ and\
  \citenamefont {Nori}}]{zhong2013fisher}%
  \BibitemOpen
  \bibfield  {author} {\bibinfo {author} {\bibfnamefont {W.}~\bibnamefont
  {Zhong}}, \bibinfo {author} {\bibfnamefont {Z.}~\bibnamefont {Sun}}, \bibinfo
  {author} {\bibfnamefont {J.}~\bibnamefont {Ma}}, \bibinfo {author}
  {\bibfnamefont {X.}~\bibnamefont {Wang}}, \ and\ \bibinfo {author}
  {\bibfnamefont {F.}~\bibnamefont {Nori}},\ }\bibfield  {title} {\emph
  {\bibinfo {title} {Fisher information under decoherence in bloch
  representation},\ }}\href {\doibase 10.1103/PhysRevA.87.022337} {\bibfield
  {journal} {\bibinfo  {journal} {Phys. Rev. A}\ }\textbf {\bibinfo {volume}
  {87}},\ \bibinfo {pages} {022337} (\bibinfo {year} {2013})}\BibitemShut
  {NoStop}%
\bibitem [{\citenamefont {Giovannetti}\ \emph {et~al.}(2004)\citenamefont
  {Giovannetti}, \citenamefont {Lloyd},\ and\ \citenamefont
  {Maccone}}]{giovannetti2004quantum}%
  \BibitemOpen
  \bibfield  {author} {\bibinfo {author} {\bibfnamefont {V.}~\bibnamefont
  {Giovannetti}}, \bibinfo {author} {\bibfnamefont {S.}~\bibnamefont {Lloyd}},
  \ and\ \bibinfo {author} {\bibfnamefont {L.}~\bibnamefont {Maccone}},\
  }\bibfield  {title} {\emph {\bibinfo {title} {Quantum-enhanced measurements:
  beating the standard quantum limit},\ }}\href {\doibase
  10.1126/science.1104149} {\bibfield  {journal} {\bibinfo  {journal}
  {Science}\ }\textbf {\bibinfo {volume} {306}},\ \bibinfo {pages} {1330}
  (\bibinfo {year} {2004})}\BibitemShut {NoStop}%
\bibitem [{\citenamefont {Kay}(1993)}]{kay1993fundamentals}%
  \BibitemOpen
  \bibfield  {author} {\bibinfo {author} {\bibfnamefont {S.~M.}\ \bibnamefont
  {Kay}},\ }\href@noop {} {\emph {\bibinfo {title} {Fundamentals of
  {Statistical Signal Processing: Estimation Theory}}}}\ (\bibinfo  {publisher}
  {Prentice-Hall, New York},\ \bibinfo {year} {1993})\BibitemShut {NoStop}%
\bibitem [{\citenamefont {Ludlow}\ \emph {et~al.}(2015)\citenamefont {Ludlow},
  \citenamefont {Boyd}, \citenamefont {Ye}, \citenamefont {Peik},\ and\
  \citenamefont {Schmidt}}]{Ludlow2015optical}%
  \BibitemOpen
  \bibfield  {author} {\bibinfo {author} {\bibfnamefont {A.~D.}\ \bibnamefont
  {Ludlow}}, \bibinfo {author} {\bibfnamefont {M.~M.}\ \bibnamefont {Boyd}},
  \bibinfo {author} {\bibfnamefont {J.}~\bibnamefont {Ye}}, \bibinfo {author}
  {\bibfnamefont {E.}~\bibnamefont {Peik}}, \ and\ \bibinfo {author}
  {\bibfnamefont {P.~O.}\ \bibnamefont {Schmidt}},\ }\bibfield  {title} {\emph
  {\bibinfo {title} {Optical atomic clocks},\ }}\href {\doibase
  10.1103/RevModPhys.87.637} {\bibfield  {journal} {\bibinfo  {journal} {Rev.
  Mod. Phys.}\ }\textbf {\bibinfo {volume} {87}},\ \bibinfo {pages} {637}
  (\bibinfo {year} {2015})}\BibitemShut {NoStop}%
\bibitem [{\citenamefont {Katori}(2011)}]{Katori2011optical}%
  \BibitemOpen
  \bibfield  {author} {\bibinfo {author} {\bibfnamefont {H.}~\bibnamefont
  {Katori}},\ }\bibfield  {title} {\emph {\bibinfo {title} {Optical lattice
  clocks and quantum metrology},\ }}\href {\doibase
  https://doi.org/10.1038/nphoton.2011.45} {\bibfield  {journal} {\bibinfo
  {journal} {Nat. Photonics}\ }\textbf {\bibinfo {volume} {5}},\ \bibinfo
  {pages} {203} (\bibinfo {year} {2011})}\BibitemShut {NoStop}%
\bibitem [{\citenamefont {Caves}(1981)}]{Caves1981quantum}%
  \BibitemOpen
  \bibfield  {author} {\bibinfo {author} {\bibfnamefont {C.~M.}\ \bibnamefont
  {Caves}},\ }\bibfield  {title} {\emph {\bibinfo {title} {Quantum-mechanical
  noise in an interferometer},\ }}\href {\doibase 10.1103/PhysRevD.23.1693}
  {\bibfield  {journal} {\bibinfo  {journal} {Phys. Rev. D}\ }\textbf {\bibinfo
  {volume} {23}},\ \bibinfo {pages} {1693} (\bibinfo {year}
  {1981})}\BibitemShut {NoStop}%
\bibitem [{\citenamefont {Taylor}\ and\ \citenamefont
  {Bowen}(2016)}]{Taylor2016quantum}%
  \BibitemOpen
  \bibfield  {author} {\bibinfo {author} {\bibfnamefont {M.~A.}\ \bibnamefont
  {Taylor}}\ and\ \bibinfo {author} {\bibfnamefont {W.~P.}\ \bibnamefont
  {Bowen}},\ }\bibfield  {title} {\emph {\bibinfo {title} {Quantum metrology
  and its application in biology},\ }}\href {\doibase
  https://doi.org/10.1016/j.physrep.2015.12.002} {\bibfield  {journal}
  {\bibinfo  {journal} {Phys. Rep.}\ }\textbf {\bibinfo {volume} {615}},\
  \bibinfo {pages} {1} (\bibinfo {year} {2016})}\BibitemShut {NoStop}%
\bibitem [{\citenamefont {Mauranyapin}\ \emph {et~al.}(2017)\citenamefont
  {Mauranyapin}, \citenamefont {Madsen}, \citenamefont {Taylor}, \citenamefont
  {Waleed},\ and\ \citenamefont {Bowen}}]{Mauranyapin2017evanescent}%
  \BibitemOpen
  \bibfield  {author} {\bibinfo {author} {\bibfnamefont {N.}~\bibnamefont
  {Mauranyapin}}, \bibinfo {author} {\bibfnamefont {L.}~\bibnamefont {Madsen}},
  \bibinfo {author} {\bibfnamefont {M.}~\bibnamefont {Taylor}}, \bibinfo
  {author} {\bibfnamefont {M.}~\bibnamefont {Waleed}}, \ and\ \bibinfo {author}
  {\bibfnamefont {W.}~\bibnamefont {Bowen}},\ }\bibfield  {title} {\emph
  {\bibinfo {title} {Evanescent single-molecule biosensing with quantum-limited
  precision},\ }}\href {\doibase https://doi.org/10.1038/nphoton.2017.99}
  {\bibfield  {journal} {\bibinfo  {journal} {Nat. Photonics}\ }\textbf
  {\bibinfo {volume} {11}},\ \bibinfo {pages} {477} (\bibinfo {year}
  {2017})}\BibitemShut {NoStop}%
\bibitem [{\citenamefont {Jones}\ \emph {et~al.}(2009)\citenamefont {Jones},
  \citenamefont {Karlen}, \citenamefont {Fitzsimons}, \citenamefont {Ardavan},
  \citenamefont {Benjamin}, \citenamefont {Briggs},\ and\ \citenamefont
  {Morton}}]{Jones2009magnetic}%
  \BibitemOpen
  \bibfield  {author} {\bibinfo {author} {\bibfnamefont {J.~A.}\ \bibnamefont
  {Jones}}, \bibinfo {author} {\bibfnamefont {S.~D.}\ \bibnamefont {Karlen}},
  \bibinfo {author} {\bibfnamefont {J.}~\bibnamefont {Fitzsimons}}, \bibinfo
  {author} {\bibfnamefont {A.}~\bibnamefont {Ardavan}}, \bibinfo {author}
  {\bibfnamefont {S.~C.}\ \bibnamefont {Benjamin}}, \bibinfo {author}
  {\bibfnamefont {G.~A.~D.}\ \bibnamefont {Briggs}}, \ and\ \bibinfo {author}
  {\bibfnamefont {J.~J.}\ \bibnamefont {Morton}},\ }\bibfield  {title} {\emph
  {\bibinfo {title} {Magnetic field sensing beyond the standard quantum limit
  using 10-spin noon states},\ }}\href {\doibase
  https://www.science.org/doi/10.1126/science.1170730} {\bibfield  {journal}
  {\bibinfo  {journal} {Science}\ }\textbf {\bibinfo {volume} {324}},\ \bibinfo
  {pages} {1166} (\bibinfo {year} {2009})}\BibitemShut {NoStop}%
\bibitem [{\citenamefont {Song}\ \emph {et~al.}(2019)\citenamefont {Song},
  \citenamefont {Xu}, \citenamefont {Li}, \citenamefont {Zhang}, \citenamefont
  {Zhang}, \citenamefont {Liu}, \citenamefont {Guo}, \citenamefont {Wang},
  \citenamefont {Ren}, \citenamefont {Hao} \emph
  {et~al.}}]{song2019generation}%
  \BibitemOpen
  \bibfield  {author} {\bibinfo {author} {\bibfnamefont {C.}~\bibnamefont
  {Song}}, \bibinfo {author} {\bibfnamefont {K.}~\bibnamefont {Xu}}, \bibinfo
  {author} {\bibfnamefont {H.-K.}\ \bibnamefont {Li}}, \bibinfo {author}
  {\bibfnamefont {Y.-R.}\ \bibnamefont {Zhang}}, \bibinfo {author}
  {\bibfnamefont {X.}~\bibnamefont {Zhang}}, \bibinfo {author} {\bibfnamefont
  {W.-X.}\ \bibnamefont {Liu}}, \bibinfo {author} {\bibfnamefont {Q.-J.}\
  \bibnamefont {Guo}}, \bibinfo {author} {\bibfnamefont {Z.}~\bibnamefont
  {Wang}}, \bibinfo {author} {\bibfnamefont {W.-H.}\ \bibnamefont {Ren}},
  \bibinfo {author} {\bibfnamefont {J.}~\bibnamefont {Hao}},  \emph {et~al.},\
  }\bibfield  {title} {\emph {\bibinfo {title} {Generation of multicomponent
  atomic {S}chr{\"o}dinger cat states of up to 20 qubits},\ }}\href {\doibase
  https://doi.org/10.1126/science.aay0600} {\bibfield  {journal} {\bibinfo
  {journal} {Science}\ }\textbf {\bibinfo {volume} {365}},\ \bibinfo {pages}
  {574} (\bibinfo {year} {2019})}\BibitemShut {NoStop}%
\bibitem [{\citenamefont {Choi}\ \emph {et~al.}(2014)\citenamefont {Choi},
  \citenamefont {Debnath}, \citenamefont {Manning}, \citenamefont {Figgatt},
  \citenamefont {Gong}, \citenamefont {Duan},\ and\ \citenamefont
  {Monroe}}]{choi2014optimal}%
  \BibitemOpen
  \bibfield  {author} {\bibinfo {author} {\bibfnamefont {T.}~\bibnamefont
  {Choi}}, \bibinfo {author} {\bibfnamefont {S.}~\bibnamefont {Debnath}},
  \bibinfo {author} {\bibfnamefont {T.~A.}\ \bibnamefont {Manning}}, \bibinfo
  {author} {\bibfnamefont {C.}~\bibnamefont {Figgatt}}, \bibinfo {author}
  {\bibfnamefont {Z.-X.}\ \bibnamefont {Gong}}, \bibinfo {author}
  {\bibfnamefont {L.-M.}\ \bibnamefont {Duan}}, \ and\ \bibinfo {author}
  {\bibfnamefont {C.}~\bibnamefont {Monroe}},\ }\bibfield  {title} {\emph
  {\bibinfo {title} {Optimal quantum control of multimode couplings between
  trapped ion qubits for scalable entanglement},\ }}\href {\doibase
  10.1103/PhysRevLett.112.190502} {\bibfield  {journal} {\bibinfo  {journal}
  {Phys. Rev. Lett.}\ }\textbf {\bibinfo {volume} {112}},\ \bibinfo {pages}
  {190502} (\bibinfo {year} {2014})}\BibitemShut {NoStop}%
\bibitem [{\citenamefont {Barends}\ \emph {et~al.}(2014)\citenamefont
  {Barends}, \citenamefont {Kelly}, \citenamefont {Megrant}, \citenamefont
  {Veitia}, \citenamefont {Sank}, \citenamefont {Jeffrey}, \citenamefont
  {White}, \citenamefont {Mutus}, \citenamefont {Fowler}, \citenamefont
  {Campbell} \emph {et~al.}}]{barends2014superconducting}%
  \BibitemOpen
  \bibfield  {author} {\bibinfo {author} {\bibfnamefont {R.}~\bibnamefont
  {Barends}}, \bibinfo {author} {\bibfnamefont {J.}~\bibnamefont {Kelly}},
  \bibinfo {author} {\bibfnamefont {A.}~\bibnamefont {Megrant}}, \bibinfo
  {author} {\bibfnamefont {A.}~\bibnamefont {Veitia}}, \bibinfo {author}
  {\bibfnamefont {D.}~\bibnamefont {Sank}}, \bibinfo {author} {\bibfnamefont
  {E.}~\bibnamefont {Jeffrey}}, \bibinfo {author} {\bibfnamefont {T.~C.}\
  \bibnamefont {White}}, \bibinfo {author} {\bibfnamefont {J.}~\bibnamefont
  {Mutus}}, \bibinfo {author} {\bibfnamefont {A.~G.}\ \bibnamefont {Fowler}},
  \bibinfo {author} {\bibfnamefont {B.}~\bibnamefont {Campbell}},  \emph
  {et~al.},\ }\bibfield  {title} {\emph {\bibinfo {title} {Superconducting
  quantum circuits at the surface code threshold for fault tolerance},\ }}\href
  {\doibase 10.1038/nature13171} {\bibfield  {journal} {\bibinfo  {journal}
  {Nature}\ }\textbf {\bibinfo {volume} {508}},\ \bibinfo {pages} {500}
  (\bibinfo {year} {2014})}\BibitemShut {NoStop}%
\bibitem [{\citenamefont {Kaufmann}\ \emph {et~al.}(2017)\citenamefont
  {Kaufmann}, \citenamefont {Ruster}, \citenamefont {Schmiegelow},
  \citenamefont {Luda}, \citenamefont {Kaushal}, \citenamefont {Schulz},
  \citenamefont {Von~Lindenfels}, \citenamefont {Schmidt-Kaler},\ and\
  \citenamefont {Poschinger}}]{kaufmann2017scalable}%
  \BibitemOpen
  \bibfield  {author} {\bibinfo {author} {\bibfnamefont {H.}~\bibnamefont
  {Kaufmann}}, \bibinfo {author} {\bibfnamefont {T.}~\bibnamefont {Ruster}},
  \bibinfo {author} {\bibfnamefont {C.~T.}\ \bibnamefont {Schmiegelow}},
  \bibinfo {author} {\bibfnamefont {M.~A.}\ \bibnamefont {Luda}}, \bibinfo
  {author} {\bibfnamefont {V.}~\bibnamefont {Kaushal}}, \bibinfo {author}
  {\bibfnamefont {J.}~\bibnamefont {Schulz}}, \bibinfo {author} {\bibfnamefont
  {D.}~\bibnamefont {Von~Lindenfels}}, \bibinfo {author} {\bibfnamefont
  {F.}~\bibnamefont {Schmidt-Kaler}}, \ and\ \bibinfo {author} {\bibfnamefont
  {U.}~\bibnamefont {Poschinger}},\ }\bibfield  {title} {\emph {\bibinfo
  {title} {Scalable creation of long-lived multipartite entanglement},\ }}\href
  {\doibase 10.1103/PhysRevLett.119.150503} {\bibfield  {journal} {\bibinfo
  {journal} {Phys. Rev. lett.}\ }\textbf {\bibinfo {volume} {119}},\ \bibinfo
  {pages} {150503} (\bibinfo {year} {2017})}\BibitemShut {NoStop}%
\bibitem [{\citenamefont {Kitagawa}\ and\ \citenamefont
  {Ueda}(1993)}]{kitagawa1993squeezed}%
  \BibitemOpen
  \bibfield  {author} {\bibinfo {author} {\bibfnamefont {M.}~\bibnamefont
  {Kitagawa}}\ and\ \bibinfo {author} {\bibfnamefont {M.}~\bibnamefont
  {Ueda}},\ }\bibfield  {title} {\emph {\bibinfo {title} {Squeezed spin
  states},\ }}\href {\doibase 10.1103/PhysRevA.47.5138} {\bibfield  {journal}
  {\bibinfo  {journal} {Phys. Rev. A}\ }\textbf {\bibinfo {volume} {47}},\
  \bibinfo {pages} {5138} (\bibinfo {year} {1993})}\BibitemShut {NoStop}%
\bibitem [{\citenamefont {Chalopin}\ \emph {et~al.}(2018)\citenamefont
  {Chalopin}, \citenamefont {Bouazza}, \citenamefont {Evrard}, \citenamefont
  {Makhalov}, \citenamefont {Dreon}, \citenamefont {Dalibard}, \citenamefont
  {Sidorenkov},\ and\ \citenamefont {Nascimbene}}]{chalopin2018quantum}%
  \BibitemOpen
  \bibfield  {author} {\bibinfo {author} {\bibfnamefont {T.}~\bibnamefont
  {Chalopin}}, \bibinfo {author} {\bibfnamefont {C.}~\bibnamefont {Bouazza}},
  \bibinfo {author} {\bibfnamefont {A.}~\bibnamefont {Evrard}}, \bibinfo
  {author} {\bibfnamefont {V.}~\bibnamefont {Makhalov}}, \bibinfo {author}
  {\bibfnamefont {D.}~\bibnamefont {Dreon}}, \bibinfo {author} {\bibfnamefont
  {J.}~\bibnamefont {Dalibard}}, \bibinfo {author} {\bibfnamefont {L.~A.}\
  \bibnamefont {Sidorenkov}}, \ and\ \bibinfo {author} {\bibfnamefont
  {S.}~\bibnamefont {Nascimbene}},\ }\bibfield  {title} {\emph {\bibinfo
  {title} {Quantum-enhanced sensing using non-classical spin states of a highly
  magnetic atom},\ }}\href {\doibase
  https://doi.org/10.1038/s41467-018-07433-1} {\bibfield  {journal} {\bibinfo
  {journal} {Nat. commun.}\ }\textbf {\bibinfo {volume} {9}},\ \bibinfo {pages}
  {4955} (\bibinfo {year} {2018})}\BibitemShut {NoStop}%
\bibitem [{\citenamefont {S{\o}rensen}\ \emph {et~al.}(2001)\citenamefont
  {S{\o}rensen}, \citenamefont {Duan}, \citenamefont {Cirac},\ and\
  \citenamefont {Zoller}}]{sorensen2001many}%
  \BibitemOpen
  \bibfield  {author} {\bibinfo {author} {\bibfnamefont {A.}~\bibnamefont
  {S{\o}rensen}}, \bibinfo {author} {\bibfnamefont {L.-M.}\ \bibnamefont
  {Duan}}, \bibinfo {author} {\bibfnamefont {J.~I.}\ \bibnamefont {Cirac}}, \
  and\ \bibinfo {author} {\bibfnamefont {P.}~\bibnamefont {Zoller}},\
  }\bibfield  {title} {\emph {\bibinfo {title} {Many-particle entanglement with
  {Bose-Einstein} condensates},\ }}\href {\doibase
  https://doi.org/10.1038/35051038} {\bibfield  {journal} {\bibinfo  {journal}
  {Nature}\ }\textbf {\bibinfo {volume} {409}},\ \bibinfo {pages} {63}
  (\bibinfo {year} {2001})}\BibitemShut {NoStop}%
\bibitem [{\citenamefont {Pezz\'e}\ and\ \citenamefont
  {Smerzi}(2009)}]{pezze2009entanglement}%
  \BibitemOpen
  \bibfield  {author} {\bibinfo {author} {\bibfnamefont {L.}~\bibnamefont
  {Pezz\'e}}\ and\ \bibinfo {author} {\bibfnamefont {A.}~\bibnamefont
  {Smerzi}},\ }\bibfield  {title} {\emph {\bibinfo {title} {Entanglement,
  nonlinear dynamics, and the {H}eisenberg limit},\ }}\href {\doibase
  10.1103/PhysRevLett.102.100401} {\bibfield  {journal} {\bibinfo  {journal}
  {Phys. Rev. Lett.}\ }\textbf {\bibinfo {volume} {102}},\ \bibinfo {pages}
  {100401} (\bibinfo {year} {2009})}\BibitemShut {NoStop}%
\bibitem [{\citenamefont {Agarwal}\ \emph {et~al.}(1997)\citenamefont
  {Agarwal}, \citenamefont {Puri},\ and\ \citenamefont
  {Singh}}]{agarwal1997atomic}%
  \BibitemOpen
  \bibfield  {author} {\bibinfo {author} {\bibfnamefont {G.~S.}\ \bibnamefont
  {Agarwal}}, \bibinfo {author} {\bibfnamefont {R.~R.}\ \bibnamefont {Puri}}, \
  and\ \bibinfo {author} {\bibfnamefont {R.~P.}\ \bibnamefont {Singh}},\
  }\bibfield  {title} {\emph {\bibinfo {title} {Atomic {S}chr\"odinger cat
  states},\ }}\href {\doibase 10.1103/PhysRevA.56.2249} {\bibfield  {journal}
  {\bibinfo  {journal} {Phys. Rev. A}\ }\textbf {\bibinfo {volume} {56}},\
  \bibinfo {pages} {2249} (\bibinfo {year} {1997})}\BibitemShut {NoStop}%
\bibitem [{\citenamefont {Leibfried}\ \emph {et~al.}(2005)\citenamefont
  {Leibfried}, \citenamefont {Knill}, \citenamefont {Seidelin}, \citenamefont
  {Britton}, \citenamefont {Blakestad}, \citenamefont {Chiaverini},
  \citenamefont {Hume}, \citenamefont {Itano}, \citenamefont {Jost},
  \citenamefont {Langer} \emph {et~al.}}]{leibfried2005creation}%
  \BibitemOpen
  \bibfield  {author} {\bibinfo {author} {\bibfnamefont {D.}~\bibnamefont
  {Leibfried}}, \bibinfo {author} {\bibfnamefont {E.}~\bibnamefont {Knill}},
  \bibinfo {author} {\bibfnamefont {S.}~\bibnamefont {Seidelin}}, \bibinfo
  {author} {\bibfnamefont {J.}~\bibnamefont {Britton}}, \bibinfo {author}
  {\bibfnamefont {R.~B.}\ \bibnamefont {Blakestad}}, \bibinfo {author}
  {\bibfnamefont {J.}~\bibnamefont {Chiaverini}}, \bibinfo {author}
  {\bibfnamefont {D.~B.}\ \bibnamefont {Hume}}, \bibinfo {author}
  {\bibfnamefont {W.~M.}\ \bibnamefont {Itano}}, \bibinfo {author}
  {\bibfnamefont {J.~D.}\ \bibnamefont {Jost}}, \bibinfo {author}
  {\bibfnamefont {C.}~\bibnamefont {Langer}},  \emph {et~al.},\ }\bibfield
  {title} {\emph {\bibinfo {title} {Creation of a six-atom ‘{S}chr{\"o}dinger
  cat’state},\ }}\href {\doibase https://doi.org/10.1038/nature04251}
  {\bibfield  {journal} {\bibinfo  {journal} {Nature}\ }\textbf {\bibinfo
  {volume} {438}},\ \bibinfo {pages} {639} (\bibinfo {year}
  {2005})}\BibitemShut {NoStop}%
\bibitem [{\citenamefont {Alexander}\ \emph {et~al.}(2020)\citenamefont
  {Alexander}, \citenamefont {Bollinger},\ and\ \citenamefont
  {Uys}}]{alexander2020generating}%
  \BibitemOpen
  \bibfield  {author} {\bibinfo {author} {\bibfnamefont {B.}~\bibnamefont
  {Alexander}}, \bibinfo {author} {\bibfnamefont {J.~J.}\ \bibnamefont
  {Bollinger}}, \ and\ \bibinfo {author} {\bibfnamefont {H.}~\bibnamefont
  {Uys}},\ }\bibfield  {title} {\emph {\bibinfo {title} {Generating
  {Greenberger-Horne-Zeilinger} states with squeezing and postselection},\
  }}\href {\doibase 10.1103/PhysRevA.101.062303} {\bibfield  {journal}
  {\bibinfo  {journal} {Phys. Rev. A}\ }\textbf {\bibinfo {volume} {101}},\
  \bibinfo {pages} {062303} (\bibinfo {year} {2020})}\BibitemShut {NoStop}%
\bibitem [{\citenamefont {Wineland}\ \emph {et~al.}(1992)\citenamefont
  {Wineland}, \citenamefont {Bollinger}, \citenamefont {Itano}, \citenamefont
  {Moore},\ and\ \citenamefont {Heinzen}}]{wineland1992spin}%
  \BibitemOpen
  \bibfield  {author} {\bibinfo {author} {\bibfnamefont {D.~J.}\ \bibnamefont
  {Wineland}}, \bibinfo {author} {\bibfnamefont {J.~J.}\ \bibnamefont
  {Bollinger}}, \bibinfo {author} {\bibfnamefont {W.~M.}\ \bibnamefont
  {Itano}}, \bibinfo {author} {\bibfnamefont {F.~L.}\ \bibnamefont {Moore}}, \
  and\ \bibinfo {author} {\bibfnamefont {D.~J.}\ \bibnamefont {Heinzen}},\
  }\bibfield  {title} {\emph {\bibinfo {title} {Spin squeezing and reduced
  quantum noise in spectroscopy},\ }}\href {\doibase 10.1103/PhysRevA.46.R6797}
  {\bibfield  {journal} {\bibinfo  {journal} {Phys. Rev. A}\ }\textbf {\bibinfo
  {volume} {46}},\ \bibinfo {pages} {R6797} (\bibinfo {year}
  {1992})}\BibitemShut {NoStop}%
\bibitem [{\citenamefont {Gross}\ \emph {et~al.}(2010)\citenamefont {Gross},
  \citenamefont {Zibold}, \citenamefont {Nicklas}, \citenamefont {Esteve},\
  and\ \citenamefont {Oberthaler}}]{gross2010nonlinear}%
  \BibitemOpen
  \bibfield  {author} {\bibinfo {author} {\bibfnamefont {C.}~\bibnamefont
  {Gross}}, \bibinfo {author} {\bibfnamefont {T.}~\bibnamefont {Zibold}},
  \bibinfo {author} {\bibfnamefont {E.}~\bibnamefont {Nicklas}}, \bibinfo
  {author} {\bibfnamefont {J.}~\bibnamefont {Esteve}}, \ and\ \bibinfo {author}
  {\bibfnamefont {M.~K.}\ \bibnamefont {Oberthaler}},\ }\bibfield  {title}
  {\emph {\bibinfo {title} {Nonlinear atom interferometer surpasses classical
  precision limit},\ }}\href {\doibase https://doi.org/10.1038/nature08919}
  {\bibfield  {journal} {\bibinfo  {journal} {Nature}\ }\textbf {\bibinfo
  {volume} {464}},\ \bibinfo {pages} {1165} (\bibinfo {year}
  {2010})}\BibitemShut {NoStop}%
\bibitem [{\citenamefont {Riedel}\ \emph {et~al.}(2010)\citenamefont {Riedel},
  \citenamefont {B{\"o}hi}, \citenamefont {Li}, \citenamefont {H{\"a}nsch},
  \citenamefont {Sinatra},\ and\ \citenamefont {Treutlein}}]{riedel2010atom}%
  \BibitemOpen
  \bibfield  {author} {\bibinfo {author} {\bibfnamefont {M.~F.}\ \bibnamefont
  {Riedel}}, \bibinfo {author} {\bibfnamefont {P.}~\bibnamefont {B{\"o}hi}},
  \bibinfo {author} {\bibfnamefont {Y.}~\bibnamefont {Li}}, \bibinfo {author}
  {\bibfnamefont {T.~W.}\ \bibnamefont {H{\"a}nsch}}, \bibinfo {author}
  {\bibfnamefont {A.}~\bibnamefont {Sinatra}}, \ and\ \bibinfo {author}
  {\bibfnamefont {P.}~\bibnamefont {Treutlein}},\ }\bibfield  {title} {\emph
  {\bibinfo {title} {Atom-chip-based generation of entanglement for quantum
  metrology},\ }}\href {\doibase https://doi.org/10.1038/nature08988}
  {\bibfield  {journal} {\bibinfo  {journal} {Nature}\ }\textbf {\bibinfo
  {volume} {464}},\ \bibinfo {pages} {1170} (\bibinfo {year}
  {2010})}\BibitemShut {NoStop}%
\bibitem [{\citenamefont {Bohnet}\ \emph {et~al.}(2016)\citenamefont {Bohnet},
  \citenamefont {Sawyer}, \citenamefont {Britton}, \citenamefont {Wall},
  \citenamefont {Rey}, \citenamefont {Foss-Feig},\ and\ \citenamefont
  {Bollinger}}]{bohnet2016quantum}%
  \BibitemOpen
  \bibfield  {author} {\bibinfo {author} {\bibfnamefont {J.~G.}\ \bibnamefont
  {Bohnet}}, \bibinfo {author} {\bibfnamefont {B.~C.}\ \bibnamefont {Sawyer}},
  \bibinfo {author} {\bibfnamefont {J.~W.}\ \bibnamefont {Britton}}, \bibinfo
  {author} {\bibfnamefont {M.~L.}\ \bibnamefont {Wall}}, \bibinfo {author}
  {\bibfnamefont {A.~M.}\ \bibnamefont {Rey}}, \bibinfo {author} {\bibfnamefont
  {M.}~\bibnamefont {Foss-Feig}}, \ and\ \bibinfo {author} {\bibfnamefont
  {J.~J.}\ \bibnamefont {Bollinger}},\ }\bibfield  {title} {\emph {\bibinfo
  {title} {Quantum spin dynamics and entanglement generation with hundreds of
  trapped ions},\ }}\href {\doibase https://doi.org/10.1126/science.aad9958}
  {\bibfield  {journal} {\bibinfo  {journal} {Science}\ }\textbf {\bibinfo
  {volume} {352}},\ \bibinfo {pages} {1297} (\bibinfo {year}
  {2016})}\BibitemShut {NoStop}%
\bibitem [{\citenamefont {Lu}\ \emph {et~al.}(2019)\citenamefont {Lu},
  \citenamefont {Zhang}, \citenamefont {Zhang}, \citenamefont {Chen},
  \citenamefont {Shen}, \citenamefont {Zhang}, \citenamefont {Zhang},\ and\
  \citenamefont {Kim}}]{lu2019global}%
  \BibitemOpen
  \bibfield  {author} {\bibinfo {author} {\bibfnamefont {Y.}~\bibnamefont
  {Lu}}, \bibinfo {author} {\bibfnamefont {S.-N.}\ \bibnamefont {Zhang}},
  \bibinfo {author} {\bibfnamefont {K.}~\bibnamefont {Zhang}}, \bibinfo
  {author} {\bibfnamefont {W.-T.}\ \bibnamefont {Chen}}, \bibinfo {author}
  {\bibfnamefont {Y.-C.}\ \bibnamefont {Shen}}, \bibinfo {author}
  {\bibfnamefont {J.-L.}\ \bibnamefont {Zhang}}, \bibinfo {author}
  {\bibfnamefont {J.-N.}\ \bibnamefont {Zhang}}, \ and\ \bibinfo {author}
  {\bibfnamefont {K.}~\bibnamefont {Kim}},\ }\bibfield  {title} {\emph
  {\bibinfo {title} {Global entangling gates on arbitrary ion qubits},\ }}\href
  {\doibase https://doi.org/10.1038/s41586-019-1428-4} {\bibfield  {journal}
  {\bibinfo  {journal} {Nature}\ }\textbf {\bibinfo {volume} {572}},\ \bibinfo
  {pages} {363} (\bibinfo {year} {2019})}\BibitemShut {NoStop}%
\bibitem [{\citenamefont {Xu}\ \emph {et~al.}(2020)\citenamefont {Xu},
  \citenamefont {Sun}, \citenamefont {Liu}, \citenamefont {Zhang},
  \citenamefont {Li}, \citenamefont {Dong}, \citenamefont {Ren}, \citenamefont
  {Zhang}, \citenamefont {Nori}, \citenamefont {Zheng} \emph
  {et~al.}}]{xu2020probing}%
  \BibitemOpen
  \bibfield  {author} {\bibinfo {author} {\bibfnamefont {K.}~\bibnamefont
  {Xu}}, \bibinfo {author} {\bibfnamefont {Z.-H.}\ \bibnamefont {Sun}},
  \bibinfo {author} {\bibfnamefont {W.-X.}\ \bibnamefont {Liu}}, \bibinfo
  {author} {\bibfnamefont {Y.-R.}\ \bibnamefont {Zhang}}, \bibinfo {author}
  {\bibfnamefont {H.-K.}\ \bibnamefont {Li}}, \bibinfo {author} {\bibfnamefont
  {H.}~\bibnamefont {Dong}}, \bibinfo {author} {\bibfnamefont {W.-H.}\
  \bibnamefont {Ren}}, \bibinfo {author} {\bibfnamefont {P.-F.}\ \bibnamefont
  {Zhang}}, \bibinfo {author} {\bibfnamefont {F.}~\bibnamefont {Nori}},
  \bibinfo {author} {\bibfnamefont {D.-N.}\ \bibnamefont {Zheng}},  \emph
  {et~al.},\ }\bibfield  {title} {\emph {\bibinfo {title} {Probing dynamical
  phase transitions with a superconducting quantum simulator},\ }}\href
  {\doibase https://doi.org/10.1126/sciadv.aba4935} {\bibfield  {journal}
  {\bibinfo  {journal} {Sci. Adv.}\ }\textbf {\bibinfo {volume} {6}},\ \bibinfo
  {pages} {eaba4935} (\bibinfo {year} {2020})}\BibitemShut {NoStop}%
\bibitem [{\citenamefont {Zhang}\ \emph {et~al.}(2017)\citenamefont {Zhang},
  \citenamefont {Zhou}, \citenamefont {Zhou}, \citenamefont {Guo},\ and\
  \citenamefont {Zhou}}]{zhang2017cavity}%
  \BibitemOpen
  \bibfield  {author} {\bibinfo {author} {\bibfnamefont {Y.-C.}\ \bibnamefont
  {Zhang}}, \bibinfo {author} {\bibfnamefont {X.-F.}\ \bibnamefont {Zhou}},
  \bibinfo {author} {\bibfnamefont {X.-X.}\ \bibnamefont {Zhou}}, \bibinfo
  {author} {\bibfnamefont {G.-C.}\ \bibnamefont {Guo}}, \ and\ \bibinfo
  {author} {\bibfnamefont {Z.-W.}\ \bibnamefont {Zhou}},\ }\bibfield  {title}
  {\emph {\bibinfo {title} {Cavity-assisted single-mode and two-mode
  spin-squeezed states via phase-locked atom-photon coupling},\ }}\href
  {\doibase 10.1103/PhysRevLett.118.083604} {\bibfield  {journal} {\bibinfo
  {journal} {Phys. Rev. Lett.}\ }\textbf {\bibinfo {volume} {118}},\ \bibinfo
  {pages} {083604} (\bibinfo {year} {2017})}\BibitemShut {NoStop}%
\bibitem [{\citenamefont {Borregaard}\ \emph {et~al.}(2017)\citenamefont
  {Borregaard}, \citenamefont {Davis}, \citenamefont {Bentsen}, \citenamefont
  {Schleier-Smith},\ and\ \citenamefont {S{\o}rensen}}]{borregaard2017one}%
  \BibitemOpen
  \bibfield  {author} {\bibinfo {author} {\bibfnamefont {J.}~\bibnamefont
  {Borregaard}}, \bibinfo {author} {\bibfnamefont {E.}~\bibnamefont {Davis}},
  \bibinfo {author} {\bibfnamefont {G.~S.}\ \bibnamefont {Bentsen}}, \bibinfo
  {author} {\bibfnamefont {M.~H.}\ \bibnamefont {Schleier-Smith}}, \ and\
  \bibinfo {author} {\bibfnamefont {A.~S.}\ \bibnamefont {S{\o}rensen}},\
  }\bibfield  {title} {\emph {\bibinfo {title} {One-and two-axis squeezing of
  atomic ensembles in optical cavities},\ }}\href {\doibase
  10.1088/1367-2630/aa8438} {\bibfield  {journal} {\bibinfo  {journal} {New J.
  Phys.}\ }\textbf {\bibinfo {volume} {19}},\ \bibinfo {pages} {093021}
  (\bibinfo {year} {2017})}\BibitemShut {NoStop}%
\bibitem [{\citenamefont {Helmerson}\ and\ \citenamefont
  {You}(2001)}]{helmerson2001creating}%
  \BibitemOpen
  \bibfield  {author} {\bibinfo {author} {\bibfnamefont {K.}~\bibnamefont
  {Helmerson}}\ and\ \bibinfo {author} {\bibfnamefont {L.}~\bibnamefont
  {You}},\ }\bibfield  {title} {\emph {\bibinfo {title} {Creating massive
  entanglement of {Bose-Einstein} condensed atoms},\ }}\href {\doibase
  10.1103/PhysRevLett.87.170402} {\bibfield  {journal} {\bibinfo  {journal}
  {Phys. Rev. Lett.}\ }\textbf {\bibinfo {volume} {87}},\ \bibinfo {pages}
  {170402} (\bibinfo {year} {2001})}\BibitemShut {NoStop}%
\bibitem [{\citenamefont {Macr\`{\i}}\ \emph {et~al.}(2020)\citenamefont
  {Macr\`{\i}}, \citenamefont {Nori}, \citenamefont {Savasta},\ and\
  \citenamefont {Zueco}}]{macri2020spin}%
  \BibitemOpen
  \bibfield  {author} {\bibinfo {author} {\bibfnamefont {V.}~\bibnamefont
  {Macr\`{\i}}}, \bibinfo {author} {\bibfnamefont {F.}~\bibnamefont {Nori}},
  \bibinfo {author} {\bibfnamefont {S.}~\bibnamefont {Savasta}}, \ and\
  \bibinfo {author} {\bibfnamefont {D.}~\bibnamefont {Zueco}},\ }\bibfield
  {title} {\emph {\bibinfo {title} {Spin squeezing by one-photon--two-atom
  excitation processes in atomic ensembles},\ }}\href {\doibase
  10.1103/PhysRevA.101.053818} {\bibfield  {journal} {\bibinfo  {journal}
  {Phys. Rev. A}\ }\textbf {\bibinfo {volume} {101}},\ \bibinfo {pages}
  {053818} (\bibinfo {year} {2020})}\BibitemShut {NoStop}%
\bibitem [{\citenamefont {Boixo}\ \emph {et~al.}(2007)\citenamefont {Boixo},
  \citenamefont {Flammia}, \citenamefont {Caves},\ and\ \citenamefont
  {Geremia}}]{boixo2007generalized}%
  \BibitemOpen
  \bibfield  {author} {\bibinfo {author} {\bibfnamefont {S.}~\bibnamefont
  {Boixo}}, \bibinfo {author} {\bibfnamefont {S.~T.}\ \bibnamefont {Flammia}},
  \bibinfo {author} {\bibfnamefont {C.~M.}\ \bibnamefont {Caves}}, \ and\
  \bibinfo {author} {\bibfnamefont {J.}~\bibnamefont {Geremia}},\ }\bibfield
  {title} {\emph {\bibinfo {title} {Generalized limits for single-parameter
  quantum estimation},\ }}\href {\doibase 10.1103/PhysRevLett.98.090401}
  {\bibfield  {journal} {\bibinfo  {journal} {Phys. Rev. Lett.}\ }\textbf
  {\bibinfo {volume} {98}},\ \bibinfo {pages} {090401} (\bibinfo {year}
  {2007})}\BibitemShut {NoStop}%
\bibitem [{\citenamefont {Demkowicz-Dobrza\ifmmode~\acute{n}\else
  \'{n}\fi{}ski}\ and\ \citenamefont {Maccone}(2014)}]{demkowicz2014using}%
  \BibitemOpen
  \bibfield  {author} {\bibinfo {author} {\bibfnamefont {R.}~\bibnamefont
  {Demkowicz-Dobrza\ifmmode~\acute{n}\else \'{n}\fi{}ski}}\ and\ \bibinfo
  {author} {\bibfnamefont {L.}~\bibnamefont {Maccone}},\ }\bibfield  {title}
  {\emph {\bibinfo {title} {Using entanglement against noise in quantum
  metrology},\ }}\href {\doibase 10.1103/PhysRevLett.113.250801} {\bibfield
  {journal} {\bibinfo  {journal} {Phys. Rev. Lett.}\ }\textbf {\bibinfo
  {volume} {113}},\ \bibinfo {pages} {250801} (\bibinfo {year}
  {2014})}\BibitemShut {NoStop}%
\bibitem [{\citenamefont {Zhang}\ and\ \citenamefont
  {Tong}(2022)}]{zhang2022approaching}%
  \BibitemOpen
  \bibfield  {author} {\bibinfo {author} {\bibfnamefont {D.-J.}\ \bibnamefont
  {Zhang}}\ and\ \bibinfo {author} {\bibfnamefont {D.}~\bibnamefont {Tong}},\
  }\bibfield  {title} {\emph {\bibinfo {title} {Approaching
  {H}eisenberg-scalable thermometry with built-in robustness against noise},\
  }}\href {\doibase https://doi.org/10.1038/s41534-022-00588-2} {\bibfield
  {journal} {\bibinfo  {journal} {npj Quantum Information}\ }\textbf {\bibinfo
  {volume} {8}},\ \bibinfo {pages} {81} (\bibinfo {year} {2022})}\BibitemShut
  {NoStop}%
\bibitem [{\citenamefont {Fan}\ and\ \citenamefont
  {Pang}(2024)}]{fan2024achieving}%
  \BibitemOpen
  \bibfield  {author} {\bibinfo {author} {\bibfnamefont {J.}~\bibnamefont
  {Fan}}\ and\ \bibinfo {author} {\bibfnamefont {S.}~\bibnamefont {Pang}},\
  }\bibfield  {title} {\emph {\bibinfo {title} {Achieving {H}eisenberg scaling
  by probe-ancilla interaction in quantum metrology},\ }}\href {\doibase
  10.1103/PhysRevA.110.062406} {\bibfield  {journal} {\bibinfo  {journal}
  {Phys. Rev. A}\ }\textbf {\bibinfo {volume} {110}},\ \bibinfo {pages}
  {062406} (\bibinfo {year} {2024})}\BibitemShut {NoStop}%
\bibitem [{\citenamefont {Chen}\ and\ \citenamefont
  {Jing}(2024)}]{chen2024qubit}%
  \BibitemOpen
  \bibfield  {author} {\bibinfo {author} {\bibfnamefont {P.}~\bibnamefont
  {Chen}}\ and\ \bibinfo {author} {\bibfnamefont {J.}~\bibnamefont {Jing}},\
  }\bibfield  {title} {\emph {\bibinfo {title} {Qubit-assisted quantum
  metrology under a time-reversal strategy},\ }}\href {\doibase
  10.1103/PhysRevA.110.062425} {\bibfield  {journal} {\bibinfo  {journal}
  {Phys. Rev. A}\ }\textbf {\bibinfo {volume} {110}},\ \bibinfo {pages}
  {062425} (\bibinfo {year} {2024})}\BibitemShut {NoStop}%
\bibitem [{\citenamefont {Cram{\'e}r}(1999)}]{cramer1999mathematical}%
  \BibitemOpen
  \bibfield  {author} {\bibinfo {author} {\bibfnamefont {H.}~\bibnamefont
  {Cram{\'e}r}},\ }\href@noop {} {\emph {\bibinfo {title} {Mathematical
  {Methods of Statistics}}}},\ \bibinfo {series} {Princeton Mathematical
  Series}, Vol.~\bibinfo {volume} {26}\ (\bibinfo  {publisher} {Princeton
  University Press, Princeton, NJ},\ \bibinfo {year} {1999})\BibitemShut
  {NoStop}%
\bibitem [{\citenamefont {Braunstein}\ and\ \citenamefont
  {Caves}(1994)}]{braunstein1994statistical}%
  \BibitemOpen
  \bibfield  {author} {\bibinfo {author} {\bibfnamefont {S.~L.}\ \bibnamefont
  {Braunstein}}\ and\ \bibinfo {author} {\bibfnamefont {C.~M.}\ \bibnamefont
  {Caves}},\ }\bibfield  {title} {\emph {\bibinfo {title} {Statistical distance
  and the geometry of quantum states},\ }}\href {\doibase
  10.1103/PhysRevLett.72.3439} {\bibfield  {journal} {\bibinfo  {journal}
  {Phys. Rev. Lett.}\ }\textbf {\bibinfo {volume} {72}},\ \bibinfo {pages}
  {3439} (\bibinfo {year} {1994})}\BibitemShut {NoStop}%
\bibitem [{\citenamefont {Braunstein}\ \emph {et~al.}(1996)\citenamefont
  {Braunstein}, \citenamefont {Caves},\ and\ \citenamefont
  {Milburn}}]{braunstein1996generalized}%
  \BibitemOpen
  \bibfield  {author} {\bibinfo {author} {\bibfnamefont {S.~L.}\ \bibnamefont
  {Braunstein}}, \bibinfo {author} {\bibfnamefont {C.~M.}\ \bibnamefont
  {Caves}}, \ and\ \bibinfo {author} {\bibfnamefont {G.~J.}\ \bibnamefont
  {Milburn}},\ }\bibfield  {title} {\emph {\bibinfo {title} {Generalized
  uncertainty relations: theory, examples, and {L}orentz invariance},\ }}\href
  {\doibase https://doi.org/10.1006/aphy.1996.0040} {\bibfield  {journal}
  {\bibinfo  {journal} {Ann. Phys.}\ }\textbf {\bibinfo {volume} {247}},\
  \bibinfo {pages} {135} (\bibinfo {year} {1996})}\BibitemShut {NoStop}%
\bibitem [{\citenamefont {Zhang}\ \emph {et~al.}(2013)\citenamefont {Zhang},
  \citenamefont {Li}, \citenamefont {Yang},\ and\ \citenamefont
  {Jin}}]{zhang2013quantum}%
  \BibitemOpen
  \bibfield  {author} {\bibinfo {author} {\bibfnamefont {Y.-M.}\ \bibnamefont
  {Zhang}}, \bibinfo {author} {\bibfnamefont {X.-W.}\ \bibnamefont {Li}},
  \bibinfo {author} {\bibfnamefont {W.}~\bibnamefont {Yang}}, \ and\ \bibinfo
  {author} {\bibfnamefont {G.-R.}\ \bibnamefont {Jin}},\ }\bibfield  {title}
  {\emph {\bibinfo {title} {Quantum fisher information of entangled coherent
  states in the presence of photon loss},\ }}\href {\doibase
  10.1103/PhysRevA.88.043832} {\bibfield  {journal} {\bibinfo  {journal} {Phys.
  Rev. A}\ }\textbf {\bibinfo {volume} {88}},\ \bibinfo {pages} {043832}
  (\bibinfo {year} {2013})}\BibitemShut {NoStop}%
\bibitem [{\citenamefont {Liu}\ \emph {et~al.}(2014)\citenamefont {Liu},
  \citenamefont {Jing}, \citenamefont {Zhong},\ and\ \citenamefont
  {Wang}}]{liu2014quantum}%
  \BibitemOpen
  \bibfield  {author} {\bibinfo {author} {\bibfnamefont {J.}~\bibnamefont
  {Liu}}, \bibinfo {author} {\bibfnamefont {X.-X.}\ \bibnamefont {Jing}},
  \bibinfo {author} {\bibfnamefont {W.}~\bibnamefont {Zhong}}, \ and\ \bibinfo
  {author} {\bibfnamefont {X.-G.}\ \bibnamefont {Wang}},\ }\bibfield  {title}
  {\emph {\bibinfo {title} {Quantum {F}isher information for density matrices
  with arbitrary ranks},\ }}\href {\doibase 10.1088/0253-6102/61/1/08}
  {\bibfield  {journal} {\bibinfo  {journal} {Commun. Theor. Phys.}\ }\textbf
  {\bibinfo {volume} {61}},\ \bibinfo {pages} {45} (\bibinfo {year}
  {2014})}\BibitemShut {NoStop}%
\bibitem [{\citenamefont {Brody}\ and\ \citenamefont
  {Graefe}(2012)}]{brody2012mixed}%
  \BibitemOpen
  \bibfield  {author} {\bibinfo {author} {\bibfnamefont {D.~C.}\ \bibnamefont
  {Brody}}\ and\ \bibinfo {author} {\bibfnamefont {E.-M.}\ \bibnamefont
  {Graefe}},\ }\bibfield  {title} {\emph {\bibinfo {title} {Mixed-state
  evolution in the presence of gain and loss},\ }}\href {\doibase
  10.1103/PhysRevLett.109.230405} {\bibfield  {journal} {\bibinfo  {journal}
  {Phys. Rev. Lett.}\ }\textbf {\bibinfo {volume} {109}},\ \bibinfo {pages}
  {230405} (\bibinfo {year} {2012})}\BibitemShut {NoStop}%
\bibitem [{\citenamefont {Giovannetti}\ \emph {et~al.}(2006)\citenamefont
  {Giovannetti}, \citenamefont {Lloyd},\ and\ \citenamefont
  {Maccone}}]{giovannetti2006quantum}%
  \BibitemOpen
  \bibfield  {author} {\bibinfo {author} {\bibfnamefont {V.}~\bibnamefont
  {Giovannetti}}, \bibinfo {author} {\bibfnamefont {S.}~\bibnamefont {Lloyd}},
  \ and\ \bibinfo {author} {\bibfnamefont {L.}~\bibnamefont {Maccone}},\
  }\bibfield  {title} {\emph {\bibinfo {title} {Quantum metrology},\ }}\href
  {\doibase 10.1103/PhysRevLett.96.010401} {\bibfield  {journal} {\bibinfo
  {journal} {Phys. Rev. Lett.}\ }\textbf {\bibinfo {volume} {96}},\ \bibinfo
  {pages} {010401} (\bibinfo {year} {2006})}\BibitemShut {NoStop}%
\bibitem [{\citenamefont {Pang}\ and\ \citenamefont
  {Brun}(2014)}]{pang2014quantum}%
  \BibitemOpen
  \bibfield  {author} {\bibinfo {author} {\bibfnamefont {S.}~\bibnamefont
  {Pang}}\ and\ \bibinfo {author} {\bibfnamefont {T.~A.}\ \bibnamefont
  {Brun}},\ }\bibfield  {title} {\emph {\bibinfo {title} {Quantum metrology for
  a general {H}amiltonian parameter},\ }}\href {\doibase
  10.1103/PhysRevA.90.022117} {\bibfield  {journal} {\bibinfo  {journal} {Phys.
  Rev. A}\ }\textbf {\bibinfo {volume} {90}},\ \bibinfo {pages} {022117}
  (\bibinfo {year} {2014})}\BibitemShut {NoStop}%
\bibitem [{\citenamefont {Braun}\ \emph {et~al.}(2018)\citenamefont {Braun},
  \citenamefont {Adesso}, \citenamefont {Benatti}, \citenamefont {Floreanini},
  \citenamefont {Marzolino}, \citenamefont {Mitchell},\ and\ \citenamefont
  {Pirandola}}]{braun2018quantum}%
  \BibitemOpen
  \bibfield  {author} {\bibinfo {author} {\bibfnamefont {D.}~\bibnamefont
  {Braun}}, \bibinfo {author} {\bibfnamefont {G.}~\bibnamefont {Adesso}},
  \bibinfo {author} {\bibfnamefont {F.}~\bibnamefont {Benatti}}, \bibinfo
  {author} {\bibfnamefont {R.}~\bibnamefont {Floreanini}}, \bibinfo {author}
  {\bibfnamefont {U.}~\bibnamefont {Marzolino}}, \bibinfo {author}
  {\bibfnamefont {M.~W.}\ \bibnamefont {Mitchell}}, \ and\ \bibinfo {author}
  {\bibfnamefont {S.}~\bibnamefont {Pirandola}},\ }\bibfield  {title} {\emph
  {\bibinfo {title} {Quantum-enhanced measurements without entanglement},\
  }}\href {\doibase 10.1103/RevModPhys.90.035006} {\bibfield  {journal}
  {\bibinfo  {journal} {Rev. Mod. Phys.}\ }\textbf {\bibinfo {volume} {90}},\
  \bibinfo {pages} {035006} (\bibinfo {year} {2018})}\BibitemShut {NoStop}%
\bibitem [{\citenamefont {Xie}\ \emph {et~al.}(2021)\citenamefont {Xie},
  \citenamefont {Zhao}, \citenamefont {Kong}, \citenamefont {Ma}, \citenamefont
  {Wang}, \citenamefont {Ye}, \citenamefont {Yu}, \citenamefont {Yang},
  \citenamefont {Xu}, \citenamefont {Wang} \emph {et~al.}}]{xie2021beating}%
  \BibitemOpen
  \bibfield  {author} {\bibinfo {author} {\bibfnamefont {T.}~\bibnamefont
  {Xie}}, \bibinfo {author} {\bibfnamefont {Z.}~\bibnamefont {Zhao}}, \bibinfo
  {author} {\bibfnamefont {X.}~\bibnamefont {Kong}}, \bibinfo {author}
  {\bibfnamefont {W.}~\bibnamefont {Ma}}, \bibinfo {author} {\bibfnamefont
  {M.}~\bibnamefont {Wang}}, \bibinfo {author} {\bibfnamefont {X.}~\bibnamefont
  {Ye}}, \bibinfo {author} {\bibfnamefont {P.}~\bibnamefont {Yu}}, \bibinfo
  {author} {\bibfnamefont {Z.}~\bibnamefont {Yang}}, \bibinfo {author}
  {\bibfnamefont {S.}~\bibnamefont {Xu}}, \bibinfo {author} {\bibfnamefont
  {P.}~\bibnamefont {Wang}},  \emph {et~al.},\ }\bibfield  {title} {\emph
  {\bibinfo {title} {Beating the standard quantum limit under ambient
  conditions with solid-state spins},\ }}\href {\doibase
  10.1126/sciadv.abg9204} {\bibfield  {journal} {\bibinfo  {journal} {Sci.
  Adv.}\ }\textbf {\bibinfo {volume} {7}},\ \bibinfo {pages} {eabg9204}
  (\bibinfo {year} {2021})}\BibitemShut {NoStop}%
\bibitem [{\citenamefont {Lachance-Quirion}\ \emph {et~al.}(2020)\citenamefont
  {Lachance-Quirion}, \citenamefont {Wolski}, \citenamefont {Tabuchi},
  \citenamefont {Kono}, \citenamefont {Usami},\ and\ \citenamefont
  {Nakamura}}]{lachance2020entanglement}%
  \BibitemOpen
  \bibfield  {author} {\bibinfo {author} {\bibfnamefont {D.}~\bibnamefont
  {Lachance-Quirion}}, \bibinfo {author} {\bibfnamefont {S.~P.}\ \bibnamefont
  {Wolski}}, \bibinfo {author} {\bibfnamefont {Y.}~\bibnamefont {Tabuchi}},
  \bibinfo {author} {\bibfnamefont {S.}~\bibnamefont {Kono}}, \bibinfo {author}
  {\bibfnamefont {K.}~\bibnamefont {Usami}}, \ and\ \bibinfo {author}
  {\bibfnamefont {Y.}~\bibnamefont {Nakamura}},\ }\bibfield  {title} {\emph
  {\bibinfo {title} {Entanglement-based single-shot detection of a single
  magnon with a superconducting qubit},\ }}\href {\doibase
  10.1126/science.aaz9236} {\bibfield  {journal} {\bibinfo  {journal}
  {Science}\ }\textbf {\bibinfo {volume} {367}},\ \bibinfo {pages} {425}
  (\bibinfo {year} {2020})}\BibitemShut {NoStop}%
\bibitem [{\citenamefont {Pellizzari}\ \emph {et~al.}(1995)\citenamefont
  {Pellizzari}, \citenamefont {Gardiner}, \citenamefont {Cirac},\ and\
  \citenamefont {Zoller}}]{pellizzari1995decoherence}%
  \BibitemOpen
  \bibfield  {author} {\bibinfo {author} {\bibfnamefont {T.}~\bibnamefont
  {Pellizzari}}, \bibinfo {author} {\bibfnamefont {S.~A.}\ \bibnamefont
  {Gardiner}}, \bibinfo {author} {\bibfnamefont {J.~I.}\ \bibnamefont {Cirac}},
  \ and\ \bibinfo {author} {\bibfnamefont {P.}~\bibnamefont {Zoller}},\
  }\bibfield  {title} {\emph {\bibinfo {title} {Decoherence, continuous
  observation, and quantum computing: A cavity {QED} model},\ }}\href {\doibase
  10.1103/PhysRevLett.75.3788} {\bibfield  {journal} {\bibinfo  {journal}
  {Phys. Rev. Lett.}\ }\textbf {\bibinfo {volume} {75}},\ \bibinfo {pages}
  {3788} (\bibinfo {year} {1995})}\BibitemShut {NoStop}%
\bibitem [{\citenamefont {Pyrkov}\ and\ \citenamefont
  {Byrnes}(2013)}]{pyrkov2013entanglement}%
  \BibitemOpen
  \bibfield  {author} {\bibinfo {author} {\bibfnamefont {A.~N.}\ \bibnamefont
  {Pyrkov}}\ and\ \bibinfo {author} {\bibfnamefont {T.}~\bibnamefont
  {Byrnes}},\ }\bibfield  {title} {\emph {\bibinfo {title} {Entanglement
  generation in quantum networks of {Bose-Einstein} condensates},\ }}\href
  {https://iopscience.iop.org/article/10.1088/1367-2630/15/9/093019} {\bibfield
   {journal} {\bibinfo  {journal} {New J. Phys.}\ }\textbf {\bibinfo {volume}
  {15}},\ \bibinfo {pages} {093019} (\bibinfo {year} {2013})}\BibitemShut
  {NoStop}%
\bibitem [{\citenamefont {Reed}\ \emph {et~al.}(2010)\citenamefont {Reed},
  \citenamefont {DiCarlo}, \citenamefont {Johnson}, \citenamefont {Sun},
  \citenamefont {Schuster}, \citenamefont {Frunzio},\ and\ \citenamefont
  {Schoelkopf}}]{reed2010high}%
  \BibitemOpen
  \bibfield  {author} {\bibinfo {author} {\bibfnamefont {M.~D.}\ \bibnamefont
  {Reed}}, \bibinfo {author} {\bibfnamefont {L.}~\bibnamefont {DiCarlo}},
  \bibinfo {author} {\bibfnamefont {B.~R.}\ \bibnamefont {Johnson}}, \bibinfo
  {author} {\bibfnamefont {L.}~\bibnamefont {Sun}}, \bibinfo {author}
  {\bibfnamefont {D.~I.}\ \bibnamefont {Schuster}}, \bibinfo {author}
  {\bibfnamefont {L.}~\bibnamefont {Frunzio}}, \ and\ \bibinfo {author}
  {\bibfnamefont {R.~J.}\ \bibnamefont {Schoelkopf}},\ }\bibfield  {title}
  {\emph {\bibinfo {title} {High-fidelity readout in circuit quantum
  electrodynamics using the {Jaynes-Cummings} nonlinearity},\ }}\href {\doibase
  10.1103/PhysRevLett.105.173601} {\bibfield  {journal} {\bibinfo  {journal}
  {Phys. Rev. Lett.}\ }\textbf {\bibinfo {volume} {105}},\ \bibinfo {pages}
  {173601} (\bibinfo {year} {2010})}\BibitemShut {NoStop}%
\bibitem [{\citenamefont {Meyer}\ \emph {et~al.}(2001)\citenamefont {Meyer},
  \citenamefont {Rowe}, \citenamefont {Kielpinski}, \citenamefont {Sackett},
  \citenamefont {Itano}, \citenamefont {Monroe},\ and\ \citenamefont
  {Wineland}}]{meyer2001experimental}%
  \BibitemOpen
  \bibfield  {author} {\bibinfo {author} {\bibfnamefont {V.}~\bibnamefont
  {Meyer}}, \bibinfo {author} {\bibfnamefont {M.~A.}\ \bibnamefont {Rowe}},
  \bibinfo {author} {\bibfnamefont {D.}~\bibnamefont {Kielpinski}}, \bibinfo
  {author} {\bibfnamefont {C.~A.}\ \bibnamefont {Sackett}}, \bibinfo {author}
  {\bibfnamefont {W.~M.}\ \bibnamefont {Itano}}, \bibinfo {author}
  {\bibfnamefont {C.}~\bibnamefont {Monroe}}, \ and\ \bibinfo {author}
  {\bibfnamefont {D.~J.}\ \bibnamefont {Wineland}},\ }\bibfield  {title} {\emph
  {\bibinfo {title} {Experimental demonstration of entanglement-enhanced
  rotation angle estimation using trapped ions},\ }}\href {\doibase
  10.1103/PhysRevLett.86.5870} {\bibfield  {journal} {\bibinfo  {journal}
  {Phys. Rev. Lett.}\ }\textbf {\bibinfo {volume} {86}},\ \bibinfo {pages}
  {5870} (\bibinfo {year} {2001})}\BibitemShut {NoStop}%
\bibitem [{\citenamefont {Ockeloen}\ \emph {et~al.}(2013)\citenamefont
  {Ockeloen}, \citenamefont {Schmied}, \citenamefont {Riedel},\ and\
  \citenamefont {Treutlein}}]{ockeloen2013quantum}%
  \BibitemOpen
  \bibfield  {author} {\bibinfo {author} {\bibfnamefont {C.~F.}\ \bibnamefont
  {Ockeloen}}, \bibinfo {author} {\bibfnamefont {R.}~\bibnamefont {Schmied}},
  \bibinfo {author} {\bibfnamefont {M.~F.}\ \bibnamefont {Riedel}}, \ and\
  \bibinfo {author} {\bibfnamefont {P.}~\bibnamefont {Treutlein}},\ }\bibfield
  {title} {\emph {\bibinfo {title} {Quantum metrology with a scanning probe
  atom interferometer},\ }}\href {\doibase 10.1103/PhysRevLett.111.143001}
  {\bibfield  {journal} {\bibinfo  {journal} {Phys. Rev. Lett.}\ }\textbf
  {\bibinfo {volume} {111}},\ \bibinfo {pages} {143001} (\bibinfo {year}
  {2013})}\BibitemShut {NoStop}%
\bibitem [{\citenamefont {Giovannetti}\ \emph {et~al.}(2011)\citenamefont
  {Giovannetti}, \citenamefont {Lloyd},\ and\ \citenamefont
  {Maccone}}]{giovannetti2011advances}%
  \BibitemOpen
  \bibfield  {author} {\bibinfo {author} {\bibfnamefont {V.}~\bibnamefont
  {Giovannetti}}, \bibinfo {author} {\bibfnamefont {S.}~\bibnamefont {Lloyd}},
  \ and\ \bibinfo {author} {\bibfnamefont {L.}~\bibnamefont {Maccone}},\
  }\bibfield  {title} {\emph {\bibinfo {title} {Advances in quantum
  metrology},\ }}\href {\doibase https://doi.org/10.1038/nphoton.2011.35}
  {\bibfield  {journal} {\bibinfo  {journal} {Nat. Photonics}\ }\textbf
  {\bibinfo {volume} {5}},\ \bibinfo {pages} {222} (\bibinfo {year}
  {2011})}\BibitemShut {NoStop}%
\bibitem [{\citenamefont {T{\'o}th}\ and\ \citenamefont
  {Apellaniz}(2014)}]{toth2014quantum}%
  \BibitemOpen
  \bibfield  {author} {\bibinfo {author} {\bibfnamefont {G.}~\bibnamefont
  {T{\'o}th}}\ and\ \bibinfo {author} {\bibfnamefont {I.}~\bibnamefont
  {Apellaniz}},\ }\bibfield  {title} {\emph {\bibinfo {title} {Quantum
  metrology from a quantum information science perspective},\ }}\href {\doibase
  10.1088/1751-8113/47/42/424006} {\bibfield  {journal} {\bibinfo  {journal}
  {J. Phys. A: Math. Theor.}\ }\textbf {\bibinfo {volume} {47}},\ \bibinfo
  {pages} {424006} (\bibinfo {year} {2014})}\BibitemShut {NoStop}%
\bibitem [{\citenamefont {Nawrocki}(2019)}]{nawrocki2015introduction}%
  \BibitemOpen
  \bibfield  {author} {\bibinfo {author} {\bibfnamefont {W.}~\bibnamefont
  {Nawrocki}},\ }\href@noop {} {\emph {\bibinfo {title} {Introduction to
  {Quantum Metrology}}}},\ \bibinfo {edition} {2nd}\ ed.\ (\bibinfo
  {publisher} {Springer Nature},\ \bibinfo {address} {Cham},\ \bibinfo {year}
  {2019})\BibitemShut {NoStop}%
\bibitem [{\citenamefont {Polino}\ \emph {et~al.}(2020)\citenamefont {Polino},
  \citenamefont {Valeri}, \citenamefont {Spagnolo},\ and\ \citenamefont
  {Sciarrino}}]{polino2020photonic}%
  \BibitemOpen
  \bibfield  {author} {\bibinfo {author} {\bibfnamefont {E.}~\bibnamefont
  {Polino}}, \bibinfo {author} {\bibfnamefont {M.}~\bibnamefont {Valeri}},
  \bibinfo {author} {\bibfnamefont {N.}~\bibnamefont {Spagnolo}}, \ and\
  \bibinfo {author} {\bibfnamefont {F.}~\bibnamefont {Sciarrino}},\ }\bibfield
  {title} {\emph {\bibinfo {title} {Photonic quantum metrology},\ }}\href
  {https://doi.org/10.1116/5.0007577} {\bibfield  {journal} {\bibinfo
  {journal} {AVS Quantum Sci.}\ }\textbf {\bibinfo {volume} {2}},\ \bibinfo
  {pages} {024703} (\bibinfo {year} {2020})}\BibitemShut {NoStop}%
\bibitem [{\citenamefont {Tabuchi}\ \emph {et~al.}(2015)\citenamefont
  {Tabuchi}, \citenamefont {Ishino}, \citenamefont {Noguchi}, \citenamefont
  {Ishikawa}, \citenamefont {Yamazaki}, \citenamefont {Usami},\ and\
  \citenamefont {Nakamura}}]{tabuchi2015coherent}%
  \BibitemOpen
  \bibfield  {author} {\bibinfo {author} {\bibfnamefont {Y.}~\bibnamefont
  {Tabuchi}}, \bibinfo {author} {\bibfnamefont {S.}~\bibnamefont {Ishino}},
  \bibinfo {author} {\bibfnamefont {A.}~\bibnamefont {Noguchi}}, \bibinfo
  {author} {\bibfnamefont {T.}~\bibnamefont {Ishikawa}}, \bibinfo {author}
  {\bibfnamefont {R.}~\bibnamefont {Yamazaki}}, \bibinfo {author}
  {\bibfnamefont {K.}~\bibnamefont {Usami}}, \ and\ \bibinfo {author}
  {\bibfnamefont {Y.}~\bibnamefont {Nakamura}},\ }\bibfield  {title} {\emph
  {\bibinfo {title} {Coherent coupling between a ferromagnetic magnon and a
  superconducting qubit},\ }}\href {\doibase 10.1126/science.aaa3693}
  {\bibfield  {journal} {\bibinfo  {journal} {Science}\ }\textbf {\bibinfo
  {volume} {349}},\ \bibinfo {pages} {405} (\bibinfo {year}
  {2015})}\BibitemShut {NoStop}%
\bibitem [{\citenamefont {Xu}\ \emph {et~al.}(2023)\citenamefont {Xu},
  \citenamefont {Gu}, \citenamefont {Li}, \citenamefont {Weng}, \citenamefont
  {Wang}, \citenamefont {Li}, \citenamefont {Wang}, \citenamefont {Zhu},\ and\
  \citenamefont {You}}]{xu2023quantum}%
  \BibitemOpen
  \bibfield  {author} {\bibinfo {author} {\bibfnamefont {D.}~\bibnamefont
  {Xu}}, \bibinfo {author} {\bibfnamefont {X.-K.}\ \bibnamefont {Gu}}, \bibinfo
  {author} {\bibfnamefont {H.-K.}\ \bibnamefont {Li}}, \bibinfo {author}
  {\bibfnamefont {Y.-C.}\ \bibnamefont {Weng}}, \bibinfo {author}
  {\bibfnamefont {Y.-P.}\ \bibnamefont {Wang}}, \bibinfo {author}
  {\bibfnamefont {J.}~\bibnamefont {Li}}, \bibinfo {author} {\bibfnamefont
  {H.}~\bibnamefont {Wang}}, \bibinfo {author} {\bibfnamefont {S.-Y.}\
  \bibnamefont {Zhu}}, \ and\ \bibinfo {author} {\bibfnamefont {J.~Q.}\
  \bibnamefont {You}},\ }\bibfield  {title} {\emph {\bibinfo {title} {Quantum
  control of a single magnon in a macroscopic spin system},\ }}\href {\doibase
  10.1103/PhysRevLett.130.193603} {\bibfield  {journal} {\bibinfo  {journal}
  {Phys. Rev. Lett.}\ }\textbf {\bibinfo {volume} {130}},\ \bibinfo {pages}
  {193603} (\bibinfo {year} {2023})}\BibitemShut {NoStop}%
\bibitem [{\citenamefont {Liu}\ \emph {et~al.}(2020)\citenamefont {Liu},
  \citenamefont {Yuan}, \citenamefont {Lu},\ and\ \citenamefont
  {Wang}}]{liu2020quantum}%
  \BibitemOpen
  \bibfield  {author} {\bibinfo {author} {\bibfnamefont {J.}~\bibnamefont
  {Liu}}, \bibinfo {author} {\bibfnamefont {H.-D.}\ \bibnamefont {Yuan}},
  \bibinfo {author} {\bibfnamefont {X.-M.}\ \bibnamefont {Lu}}, \ and\ \bibinfo
  {author} {\bibfnamefont {X.-G.}\ \bibnamefont {Wang}},\ }\bibfield  {title}
  {\emph {\bibinfo {title} {Quantum {F}isher information matrix and
  multiparameter estimation},\ }}\href {\doibase 10.1088/1751-8121/ab5d4d}
  {\bibfield  {journal} {\bibinfo  {journal} {J. Phys. A: Math. Theor.}\
  }\textbf {\bibinfo {volume} {53}},\ \bibinfo {pages} {023001} (\bibinfo
  {year} {2020})}\BibitemShut {NoStop}%
\bibitem [{\citenamefont {Tan}\ \emph {et~al.}(2021)\citenamefont {Tan},
  \citenamefont {Narasimhachar},\ and\ \citenamefont {Regula}}]{tan2021fisher}%
  \BibitemOpen
  \bibfield  {author} {\bibinfo {author} {\bibfnamefont {K.~C.}\ \bibnamefont
  {Tan}}, \bibinfo {author} {\bibfnamefont {V.}~\bibnamefont {Narasimhachar}},
  \ and\ \bibinfo {author} {\bibfnamefont {B.}~\bibnamefont {Regula}},\
  }\bibfield  {title} {\emph {\bibinfo {title} {Fisher information universally
  identifies quantum resources},\ }}\href {\doibase
  10.1103/PhysRevLett.127.200402} {\bibfield  {journal} {\bibinfo  {journal}
  {Phys. Rev. Lett.}\ }\textbf {\bibinfo {volume} {127}},\ \bibinfo {pages}
  {200402} (\bibinfo {year} {2021})}\BibitemShut {NoStop}%
\bibitem [{\citenamefont {Kounalakis}\ \emph {et~al.}(2022)\citenamefont
  {Kounalakis}, \citenamefont {Bauer},\ and\ \citenamefont
  {Blanter}}]{kounalakis2022analog}%
  \BibitemOpen
  \bibfield  {author} {\bibinfo {author} {\bibfnamefont {M.}~\bibnamefont
  {Kounalakis}}, \bibinfo {author} {\bibfnamefont {G.~E.~W.}\ \bibnamefont
  {Bauer}}, \ and\ \bibinfo {author} {\bibfnamefont {Y.~M.}\ \bibnamefont
  {Blanter}},\ }\bibfield  {title} {\emph {\bibinfo {title} {Analog quantum
  control of magnonic cat states on a chip by a superconducting qubit},\
  }}\href {\doibase 10.1103/PhysRevLett.129.037205} {\bibfield  {journal}
  {\bibinfo  {journal} {Phys. Rev. Lett.}\ }\textbf {\bibinfo {volume} {129}},\
  \bibinfo {pages} {037205} (\bibinfo {year} {2022})}\BibitemShut {NoStop}%
\bibitem [{\citenamefont {Rameshti}\ \emph {et~al.}(2022)\citenamefont
  {Rameshti}, \citenamefont {Kusminskiy}, \citenamefont {Haigh}, \citenamefont
  {Usami}, \citenamefont {Lachance-Quirion}, \citenamefont {Nakamura},
  \citenamefont {Hu}, \citenamefont {Tang}, \citenamefont {Bauer},\ and\
  \citenamefont {Blanter}}]{rameshti2022cavity}%
  \BibitemOpen
  \bibfield  {author} {\bibinfo {author} {\bibfnamefont {B.~Z.}\ \bibnamefont
  {Rameshti}}, \bibinfo {author} {\bibfnamefont {S.~V.}\ \bibnamefont
  {Kusminskiy}}, \bibinfo {author} {\bibfnamefont {J.~A.}\ \bibnamefont
  {Haigh}}, \bibinfo {author} {\bibfnamefont {K.}~\bibnamefont {Usami}},
  \bibinfo {author} {\bibfnamefont {D.}~\bibnamefont {Lachance-Quirion}},
  \bibinfo {author} {\bibfnamefont {Y.}~\bibnamefont {Nakamura}}, \bibinfo
  {author} {\bibfnamefont {C.-M.}\ \bibnamefont {Hu}}, \bibinfo {author}
  {\bibfnamefont {H.~X.}\ \bibnamefont {Tang}}, \bibinfo {author}
  {\bibfnamefont {G.~E.}\ \bibnamefont {Bauer}}, \ and\ \bibinfo {author}
  {\bibfnamefont {Y.~M.}\ \bibnamefont {Blanter}},\ }\bibfield  {title} {\emph
  {\bibinfo {title} {Cavity magnonics},\ }}\href {\doibase
  https://doi.org/10.1016/j.physrep.2022.06.001} {\bibfield  {journal}
  {\bibinfo  {journal} {Phys. Rep.}\ }\textbf {\bibinfo {volume} {979}},\
  \bibinfo {pages} {1} (\bibinfo {year} {2022})}\BibitemShut {NoStop}%
\end{thebibliography}%

\end{document}